\documentclass[a4paper,12pt]{article}

\input cohmacros.sty

\usepackage{citeref}
\def\at#1{[*** \att #1 ***]}  
\def\at#1{} 

\advance\textheight3.2cm        
\advance\topmargin-2.4cm
\advance\textwidth2.1cm
\advance\evensidemargin-0.3cm
\advance\oddsidemargin-0.3cm

\usepackage{color}

\newcommand{\vertiii}[1]{{\left\vert\kern-0.25ex\left\vert\kern-0.25ex\left\vert #1
    \right\vert\kern-0.25ex\right\vert\kern-0.25ex\right\vert}}

\newcommand{\doma}[1]{{ \rm dom}}

\begin{document}

\vspace*{-2cm}
\begin{center}
{\LARGE \bf Introduction to coherent quantization} \\

\vspace{1cm}

\centerline{\sl {\large \bf Arnold Neumaier}}

\vspace{0.25cm}

\centerline{\sl Fakult\"at f\"ur Mathematik, Universit\"at Wien}
\centerline{\sl Oskar-Morgenstern-Platz 1, A-1090 Wien, Austria}
\centerline{\sl email: Arnold.Neumaier@univie.ac.at}
\centerline{\sl WWW: \url{http://www.mat.univie.ac.at/~neum}}

\vspace{0.5cm}

\centerline{\sl {\large \bf Arash Ghaani Farashahi}}

\vspace{0.25cm}

\centerline{\sl Department of Mechanical Engineering} 
\centerline{\sl National University of Singapore}
\centerline{\sl 9 Engineering Drive 1, Singapore 117575, Singapore}
\centerline{\sl email: arash.ghaanifarashahi@nus.edu.sg}
\centerline{\sl email: ghaanifarashahi@outlook.com}
\centerline{\sl WWW: \url{https://sites.google.com/site/ghaanifarashahi/}}
\end{center}


February 2, 2022


\vfill
\bfi{Abstract.}
This paper studies coherent quantization, the way operators in the 
quantum space of a coherent space -- defined in the recent book 
'Coherent Quantum Mechanics' by the first author -- can be studied in 
terms of objects defined directly on the coherent space. The results 
may be viewed as a generalization of geometric quantization, including 
the non-unitary case.

Care has been taken to work with the weakest meaningful topology and
to assume as little as possible about the spaces and groups involved.
Unlike in geometric quantization, the groups are not assumed to be
compact, locally compact, or finite-dimensional. This implies that the
setting can be successfully applied to quantum field theory, where the
groups involved satisfy none of these properties.

The paper characterizes linear operators acting on the quantum space of
a coherent space in terms of their coherent matrix elements.
Coherent maps and associated symmetry groups for coherent spaces are
introduced, and formulas are derived for the quantization of coherent
maps.

The importance of coherent maps for quantum mechanics is due to the
fact that there is a quantization map that associates homomorphically 
with every coherent map a linear operator from the quantum space into 
itself. The quantization map generalizes the second quantization 
procedure for free classical fields to symmetry groups of general 
coherent spaces. Field quantization is obtained by specialization to 
Klauder spaces, whose quantum spaces are the bosonic Fock spaces. 
Implied by the new approach is a short, coordinate-free derivation of 
all basic properties of creation and annihilation operators in Fock 
spaces.




\bigskip
{\bf Acknowledgments.} Thanks to Rahel Kn\"opfel, David Bar Moshe,
and Hermann Schichl for useful discussions related to the subject, and
to a referee for many useful remarks.

\tableofcontents 

\section{Introduction}

This paper studies coherent quantization, the way operators in the 
quantum space of a coherent space -- defined in the recent book 
'Coherent Quantum Mechanics' \cite{Neu.CQP} by the first author -- 
can be studied in terms of objects defined directly on the coherent 
space. The results may be viewed as a generalization of geometric 
quantization, including the non-unitary case.

Care has been taken to work with the weakest meaningful topology and
to assume as little as possible about the groups involved.
In particular, unlike in geometric quantization
(\sca{Bates \& Weinstein} \cite{BatW}, \sca{Woodhouse} \cite{Woo}),
we do not assume the groups to be compact, locally compact, or
finite-dimensional. This implies that the setting can be successfully
applied to quantum field theory, where the groups involved satisfy none
of these properties.

More specifically, we characterize linear operators acting on the
quantum space of a coherent space in terms of their coherent
matrix elements. We discuss coherent maps and associated symmetry
groups for the coherent spaces introduced in \sca{Neumaier}
\cite{Neu.CQP}, and derive formulas for the quantization of
coherent maps.

An early paper by \sca{It\^o} \cite{Ito} describes unitary group
representations in terms of what are now called (generalized) coherent
states.
Group theoretic work on the subject was greatly extended by
\sca{Perelomov} \cite{Per.cs,Per}, \sca{Gilmore} \cite{Gil.cs}, and
others; see, e.g., the survey by \sca{Zhang} et al. \cite{ZhaFG}.
The present setting may be viewed as a generalization of this and of 
vector coherent states (\sca{Rowe} et al. \cite{RowRG}) to the 
non-unitary case.

The importance of coherent maps for quantum mechanics is due to the
fact proved in Theorem \ref{t.Gamma} below that there is a
\bfi{quantization map} $\Gamma$ that associates homomorphically
with every coherent map $A$ a linear operator $\Gamma(A)$ on the
augmented quantum space $\Qz^\*(Z)$ that maps the quantum space
$\Qz(Z)$ into itself. In the special case (discussed in Section 
\ref{s.Klauder} below) where $Z$ is a Klauder space, the quantum space 
$\Qz(Z)$ is a dense subspace of a bosonic Fock space and the 
quantization map is the restriction of the second quantization map of 
\sca{Derezi\'nski \& G\'erard} \cite{DerG} to $\Qz(Z)$. 
Thus the quantization map generalizes the \bfi{second quantization} 
procedure of free classical fields to symmetry groups of general 
coherent spaces. 


\bigskip
{\bf Contents.}
In the present section we review notation, terminology, and some
results on coherent spaces and their quantum spaces, introduced in 
\sca{Neumaier} \cite{Neu.CQP}, on which the present paper is based.
Section \ref{s.adFunc} provides fundamental but abstract necessary and
sufficient conditions for recognizing when a kernel, i.e., a map from
a coherent space into itself is a shadow (i.e., definable in terms of
coherent matrix elements), and hence determines an operator on the
corresponding quantum space.

Section \ref{s.cohMaps} discusses symmetries of a coherent space, one
of the most important concepts for studying and using coherent spaces.
Indeed, most of the applications of coherent spaces in quantum mechanics
and quantum field theory rely on the presence of a large symmetry group.
The main reason is that there is a quantization map that furnishes a
representation of the semigroup of coherent maps on the quantum space,
and thus provides easy access to a class of very well-behaved linear
operators on the quantum space.
Section \ref{s.hom} looks at self-mappings of coherent spaces satisfying
homogeneity or separability properties. These often give simple but
important coherent maps.
In Section \ref{s.slender} we prove quantization theorems for a
restricted class of coherent spaces for which many operators on a
quantum space have a simple description in terms of normal kernels.
These generalize the normal ordering of operators familiar from quantum
field theory.

In the final Section \ref{s.Klauder} we discuss in some detail the
coherent quantization of Klauder spaces, a class of coherent spaces
with a large semigroup of coherent maps, introduced in \sca{Neumaier} 
\cite{Neu.CQP}. The corresponding coherent 
states are closely related to those introduced by \sca{Schr\"odinger}
\cite{Schr}) and made prominent in quantum optics by \sca{Glauber}
\cite{Gla}. The quantum spaces of Klauder spaces are the bosonic Fock
spaces, which play a very important role in quantum field theory
(\sca{Baez} et al. \cite{BaeSZ}, \sca{Glimm \& Jaffe} \cite{GliJ}),
and the theory of Hida distributions in the white noise calculus
for classical stochastic processes (\sca{Hida \& Si} \cite{HidS},
\sca{Hida \& Streit} \cite{HidSt}, \sca{Obata} \cite{Oba}).
In particular, we give a coordinate-free derivation of the basic
properties of creation and annihilation operators in Fock spaces.

\subsection{Euclidean spaces}

In this paper, we use the notation and terminology of \sca{Neumaier}
\cite{Neu.CQP}, quickly reviewed here.

We write $\Cz$ for the field of complex numbers, $\Cz^\*$ for
the multiplicative group of nonzero complex numbers, and 
$\alpha^*=\ol\alpha$ for the complex conjugate of $\alpha\in\Cz$.
$\Cz^X$ denotes the vector space of all maps from a set $X$ to $\Cz$.

A (complex) \bfi{Euclidean space} is a complex vector space $\Hz$ with a
Hermitian form that assigns to $\phi,\psi\in\Hz$ the
\bfi{Hermitian inner product} $\<\phi,\psi\>\in\Cz$, antilinear in the
first and linear in the second argument, such that
\lbeq{e.her}
\ol{\<\phi,\psi\>}=\<\psi,\phi\>,
\eeq
\lbeq{e.def}
\<\psi,\psi\>>0 \Forall \psi\in\Hz\setminus\{0\}.
\eeq
Associated with $\Hz$ is the triple of spaces
\lbeq{e.HzHzx}
\Hz\subseteq \ol\Hz\subseteq \Hz^\*,
\eeq
where $\ol\Hz$ is a Hilbert space completion of $\Hz$, and $\Hz$ is
dense in the vector space $\Hz^\*$ of all antilinear
functionals on $\Hz$, with
\[
\psi(\phi):=\<\phi,\psi\> \for \phi\in\Hz.
\]
$\Hz^\*$ carries a Hermitian partial inner product $\phi^*\psi$ with
\lbeq{e.ip}
\phi^*\psi=\<\phi,\psi\>\for \phi,\psi\in\ol\Hz.
\eeq
and is a PIP space in the sense of \sca{Antoine \& Trapani} \cite{PIP}.
\gzit{e.ip} also serves as a definitioon of the linear functional 
$\phi^*$ on $\ol\Hz$ for $\phi\in\ol\Hz$.

Let $U$ and $V$ be (complex) topological vector spaces. We write
$\Lin(U,V)$ for the vector space of all continuous linear mappings from
$U$ to $V$, $\Lin U$ for $\Lin(U,U)$, and $U^\*$ for the \bfi{antidual}
of $U$, the space of all continuous antilinear mappings from $U$ to 
$\Cz$. We identify $V$ with the space $\Lin(\Cz,V)$. If $U,V$ are 
Euclidean spaces and $A\in \Lin(U,V)$, the adjoint operator 
$A^*\in\Lin(V^\*,U^\*)$ is defined by
\lbeq{e.adj}
(A^*\phi)(\psi):=\phi(A\psi)\for \phi\in V^\*,\psi\in U.
\eeq
We write $\Linx\Hz:=\Lin(\Hz,\Hz^\*)$ for the vector space of
linear operators from a Euclidean space $\Hz$ to its antidual. Note 
that the adjoint of an operator in $\Linx\Hz$ is again in $\Linx\Hz$.

As in \cite{Neu.CQP}, continuity in a Euclidean vector space $U$ is 
always understood in the finest locally convex topology, and continuity 
in its antidual $U^\*$ is always understood in the weak-* topology.
In particular, if $U,V$ are Euclidean spaces, all linear and antilinear 
functionals on $U$ and all linear mappings from $U\to V^\*$ are 
continuous.

\subsection{Coherent spaces}

The notion of a coherent space is a nonlinear version of the notion of
a complex Euclidean space: The vector space axioms are dropped while
the notion of inner product is kept.
Coherent spaces provide a setting for the study of geometry in a
different direction than traditional metric, topological, and
differential geometry. Just as it pays to study the properties of
manifolds independently of their embedding into a Euclidean space,
so it appears fruitful to study the properties of coherent spaces
independent of their embedding into a Hilbert space.

A \bfi{coherent space} is a nonempty set $Z$ with a distinguished
function $K:Z\times Z\to\Cz$ of positive type called the
\bfi{coherent product}. Thus
\lbeq{e.Kherm}
\ol{K(z,z')}=K(z',z),
\eeq
and for all $z_1,\ldots,z_n\in Z$, the $n\times n$ matrix $G$ with
entries $G_{jk}=K(z_j,z_k)$ is positive semidefinite.

The coherent space $Z$ is called \bfi{nondegenerate} if
\[
K(z'',z')=K(z,z')~~\forall~ z'\in Z \implies z'' = z.
\]
The coherent space $Z$ is called \bfi{projective} of  \bfi{degree}
$e\in\Zz\setminus\{0\}$  if there is a
\bfi{scalar multiplication} that assigns to each $\lambda\in\Cz^\*$
and each $z\in Z$ a point $\lambda z \in Z$ such that
\lbeq{e.projective}
K(z',\lambda z) = \lambda^e K(z',z).
\eeq
Equivalently,
\[
|\lambda z\>=\lambda^e|z\>.
\]
For any coherent space $Z$, the \bfi{projective extension} of $Z$ of
degree $e$ (a nonzero integer) is the coherent space
$PZ:=\Cz^\*\times Z$ with coherent product
\lbeq{e.projextensionK}
K_{\rm pe}((\lambda,z),(\lambda',z'))
~:=~ \ol{\lambda}^e\, K(z,z') \, \lambda'^e
\eeq
and scalar multiplication $\lambda'(\lambda,z):=(\lambda'\lambda,z)$,
defined in \cite[Proposition 5.8.5]{Neu.CQP}, with the same quantum
space as $Z$.

Throughout the paper, $Z$ is a fixed coherent space with coherent
product $K$. A \bfi{quantum space} $\Qz(Z)$ of $Z$ is a Euclidean space
spanned (algebraically) by a distinguished set of vectors $|z\>$
($z\in Z$) called \bfi{coherent states} satisfying
\lbeq{e.cohProd}
\<z|z'\> ~=~ K(z,z') \for z,z'\in Z,
\eeq
where $\<z|:=|z\>^*$.
The associated \bfi{augmented quantum space} $\Qz^\*(Z)$, the antidual
of $\Qz(Z)$, contains the \bfi{completed quantum space} $\ol\Qz(Z)$,
the Hilbert space completion of $\Qz(Z)$. 
By \cite[Section 4.3]{Neu.CQP},
any linear or antilinear map from a quantum space of a coherent space 
into $\Cz$ is continuous, and by the above convention for the topology 
of Euclidean spaces, any linear or antilinear map from a 
quantum space of a coherent space into its antidual is continuous, too.

\section{Quantization through admissibility conditions}\label{s.adFunc}

We regard the \bfi{quantization} of a coherent space $Z$ as the problem
of describing interesting classes of linear operators from $\Linx\Qz(Z)$
and their properties in terms of objects more tangibly defined on $Z$.
The key to coherent quantization is the observation that one can
frequently define and manipulate operators on the quantum space in
terms of their coherent matrix elements, without needing a more
explicit description in terms of differential or integral operators on
a Hilbert space of functions.

A \bfi{kernel} on $Z$ is a map $X\in \Cz^{Z\times Z}$. The \bfi{shadow} 
of a linear operator $\X\in \Linx \Qz(Z)$ is the kernel
$\sh \X\in \Cz^{Z\times Z}$ defined by
(cf. \sca{Klauder} \cite{Kla.III})
\[
\sh \X(z,z'):=\<z|\X|z'\> \for z,z'\in Z.
\]
Thus shadows represent the information in the \bfi{coherent matrix
elements} $\<z|\X|z'\>$ of an operator $\X$.

This section discusses admissibility conditions. They provide
fundamental but abstract necessary and sufficient conditions for
recognizing when a kernel is a shadow and hence determines an operator
$\X\in\Linx \Qz(Z)$. Later sections then provide applications to more
concrete situations.

The admissibility conditions are infinite generalizations of the simple
situation when $Z$ is finite. In this case we may w.l.o.g. take
$Z=\{1,2,\ldots,n\}$ and regard kernels as $n\times n$ matrices.
Then the coherent producxt is just a positive semidefinite matrix
$K=R^*R$, and the shadow of an operator $\X$ is $X=R^*\X R$.
Admissibility of $X$, here equivalent with strong admissibility, is the
condition that for any column vector $c$, $Rc=0$ implies $Xc=0 $ and
$X^*c=0$, which forces $X$ to have at most the same rank as $K$.
It is not difficult to see (and follows from the results below) that
this condition implies that $X$ has the form $X=R^*\X R$ for some
matrix $\X$, so that an admissible $X$ is indeed a shadow.

\subsection{Admissibility}

Let $Z$ be a coherent space with the coherent product $K$.
We want to characterize the functions $f:Z\to\Cz$ for which there is
an antilinear mapping $\psi:\Qz(Z)\to\Cz$ such that
\lbeq{e.zpsi1}
f(z)=\<z|\psi:=\psi(|z\>) \Forall z\in Z.
\eeq
We call a function $f:Z\to \Cz$ \bfi{admissible} if for arbitrary
finite sequences of complex numbers $c_k$ and points $z_k\in Z$,
\lbeq{e.fIffSym}
\sum c_k K(z_k,z)=0 ~~\forall\,z\in Z
\implies \sum c_k f(z_k)=0.
\eeq
For example, for $K(z,z')=0$ for all $z,z'$ one gets a trivial coherent
space whose quantum space is $\{0\}$, and only the zero function is
admissible. On the other hand, a condition guaranteeing that every map
is admissible is given in Proposition \ref{p.all} below.

\begin{thm}\label{t.adEquiv}
Let $Z$ be a coherent space. For a  quantum space $\Qz(Z)$
of $Z$, the following conditions on a function $f:Z\to \Cz$ are
equivalent.

(i) There is an antilinear functional $\psi:\Qz(Z)\to\Cz$ (i.e., a 
$\psi\in\Qz^\*(Z)$) such that \gzit{e.zpsi1} holds.

(ii) For arbitrary finite sequences of complex numbers $c_k$ and
points $z_k\in Z$,
\lbeq{e.psiIffSym1}
\sum c_k |z_k\>=0 \implies \sum \ol c_k f(z_k)=0.
\eeq
(iii) $f$ is  admissible.

Moreover, in (i), $\psi$ is uniquely determined by $f$.
\end{thm}

\bepf
(ii)$\Leftrightarrow$(i): Let $f:Z\to\Cz$ be a function satisfying
\gzit{e.psiIffSym1}. We define the antilinear functional
$\psi\in\Qz(Z)\to\Cz$ by
\lbeq{inv.s.t}
\psi\Big(\sum c_k|z_k\>\Big)
 :=\sum\ol{c_k}f(z_k)
\Forall \sum c_k|z_k\>\in\Qz(Z).
\eeq
Because of \gzit{e.psiIffSym1}, $\psi$ is well-defined; it is clearly
antilinear. Thus, $\psi$ defines an antilinear functional on the
 quantum space $\Qz(Z)$. Specializing \gzit{inv.s.t} to the
case of a sum containing a single term only gives
\[
\<z|\psi=\psi(|z\>)=f(z) \for z\in Z,
\]
so that $\psi$ satisfies \gzit{e.zpsi1}.
If \gzit{e.zpsi1} also holds for $\psi'$ in place of $\psi$ then
$\psi=\psi'$ since the coherent states span $\Qz(Z)$. This shows that
$\psi$ is uniquely determined by $f$ and \gzit{e.zpsi1}.

Conversely, let $f:Z\to\Cz$ be a function that satisfies \gzit{e.zpsi1}
for some antilinear functional $\psi$ on $\Qz(Z)$. If the left hand
 side of \gzit{e.psiIffSym1} holds then
\[
\sum \ol c_kf(z_k)=\sum\ol c_k\<z_k|\psi
=\psi\Big(\sum c_k|z_k\>\Big)=0.
\]
(iii)$\Leftrightarrow$(ii): Clearly \gzit{e.fIffSym} is equivalent to
\[
\sum \ol c_k K(z_k,z)=0 ~~\forall\,z\in Z
\implies \sum \ol c_k f(z_k)=0,
\]
The left hand side of \gzit{e.psiIffSym1} is equivalent to
\[
0=\sum \ol c_k\<z_k|z\>=\sum \ol c_k K(z_k,z') \for z\in Z.
\]
Since
\[
\<z|\sum c_k|z_k\>=\sum c_k\<z|z_k\>=\sum c_kK(z,z_k)
=\ol{\sum \ol c_kK(z_k,z)},
\]
this is equivalent to \gzit{e.psiIffSym1}.
\epf

The \bfi{admissibility space} of $Z$ is the set $\Az(Z)$ of all
admissible functions over the coherent space $Z$.
It is easy to see that $\Az(Z)$ is a vector
space with respect to pointwise addition of functions and pointwise
multiplication by complex numbers.

\begin{thm}\label{t.Theta}
Let $Z$ be a coherent space and let $\Qz(Z)$ be a  quantum
space of $Z$.

(i) For every admissible function $f:Z\to\Cz$,
\lbeq{Theta}
\theta_f\Big(\sum c_k |z_k\>\Big):=\sum\ol{c_k}f(z_k)
\for \sum c_k|z_k\>\in\Qz(Z),
\eeq
defines a continuous antilinear functional on $\Qz(Z)$.

(ii) The \bfi{identification map} $\Theta:\Az(Z)\to\Qz(Z)^\*$ given by
\lbeq{e.Theta}
\Theta(f):=\theta_f
\eeq
is a vector space isomorphism.
In particular, the admissibility space $\Az(Z)$ can be equipped with a
locally convex topology such that the linear map $\Theta$ is a 
homeomorphism.
\end{thm}

\bepf
(i) Let $f:Z\to\Cz$ be an admissible function. By Theorem
\ref{t.adEquiv}, $\theta_f=\psi$ is the unique vector in $\Qz(Z)^\*$
satisfying \gzit{e.zpsi1}.

(ii) By (i), the linear map
$\Theta:\Az(Z)\to\Qz(Z)^\*$ given by $\Theta(f):=\theta_f$ is a vector
space homomorphism. Let $\psi\in\Qz(Z)^\*$ be a given antilinear 
functional and define $f:Z\to\Cz$ via $f(z):=\<z|\psi$, for
all $z\in Z$. Then, it is easy to check that $f\in\Az(Z)$ and
$\theta_f=\psi$. Thus $\Theta$ is an isomorphism.
\epf

\begin{cor}
Let $Z$ be a coherent space. The admissible spaces $\Az(Z)$, $\Az(PZ)$,
and $\Az([Z])$ are canonically isomorphic as topological vector space.
\end{cor}

\subsection{Kernels and shadows}

For any kernel $X$ we define the related kernels $X^T$, $\ol X$, and
$X^*$ by
\[
X^T(z,z'):=X(z',z),~~~\ol X(z,z'):=\ol{X(z,z')},~~~
X^*(z,z'):=\ol{X(z',z)}.
\]
Clearly,
\[
X^{TT}=\ol{\ol X}=X^{**}=X,~~~X^*=\ol X^T=\ol{X^T}.
\]
For example, any coherent product is a kernel $K$; it is Hermitian
iff $K^T=K$. Given a kernel $X$ and $z\in Z$, we define the functions
$X(z,\cdot),X(\cdot,z)\in \Cz^Z$ by
\[
X(\cdot,z)(z'):=X(z',z),~~X(z,\cdot)(z'):=X(z,z') \for z'\in Z.
\]

\begin{prop}\label{p.shad}~\\
(i) The shadow of the identity operator $1$ is $\sh 1=K$.

(ii) For $\X\in\Linx\Qz(Z)$, the \bfi{adjoint} $|\X^*\in\Linx\Qz(Z)$,
defined by
\[
\X^*\psi(\phi):=\ol{\X\phi(\psi)},
\]
satisfies
\[
\<z|\X^*|z'\>=\ol{\<z'|\X|z\>}\Forall z,z'\in Z,
\]
\[
(\sh \X)^*=\sh {\X^*}.
\]
\end{prop}

\bepf
(i) holds since $\sh 1(z,z')=\<z|1|z'\>=\<z|z'\>=K(z,z')$ for all
$z,z'\in Z$.

(ii) Linearity implies already $\X^*\in\Linx\Qz(Z)$. Then, for
$z,z'\in Z$,
\[
\<z|\X^*|z'\>=\X^*|z'\>(|z\>)=\ol{\X|z\>(|z'\>)}=\ol{\<z'|\X|z\>},
\]
\[
 \sh{\X^*}(z,z')=\<z|\X^*|z'\>=\ol{\<z'|\X|z\>}=(\sh \X)^*(z,z').
\]
\epf

The following characterization of shadows is the fundamental theorem
on which all later quantization results are based.

\begin{thm}\label{t.opExist}
Let $Z$ be a coherent space. A kernel $X\in \Cz^{Z\times Z}$ is a 
shadow iff $X(z,\cdot)$ and $\ol{X}(\cdot,z)$ are admissible for all 
$z\in Z$.
In this case there is a unique operator $\X\in \Linx \Qz(Z)$ whose
shadow is $X$, i.e.,
\lbeq{e.zXz}
\<z|\X|z'\>=X(z,z') \for z,z'\in Z.
\eeq
\end{thm}

\bepf
Let $z\in Z$, $\X\in\Linx\Qz(Z)$, and $X:=\sh \X$. Then
$\X|z\>\in\Qz(Z)^\*$, and for $z_1,...,z_n\in Z$ and 
$c_1,...,c_n\in \Cz$, we have
\[
\ol X(z_\ell,z)=\ol{\sh \X(z_\ell,z)}=\ol{\<z_\ell|\X|z\>},
\]
hence
\[
\sum_\ell \ol c_\ell\ol X(z_\ell,z)\
=\sum_\ell \ol{c_\ell\<z_\ell|\X|z\>}
= \ol{\Big(\sum_\ell c_\ell|z_\ell\>\Big)\X|z\>}.
\]
By Theorem \ref{t.adEquiv}, 
this implies that $\ol X(\cdot,z)$ is admissible.
For $z,z'\in Z$, we have by Proposition \ref{p.shad}(ii), 
\[
X(z,z')=\<z|\X|z'\>=\ol{\<z'|\X^*|z\>}=\ol{\sh\X^*}(z',z).
\]
Hence $X(z,\cdot)=\ol{\sh\X^*}(\cdot,z)$ for all $z\in Z$.
This implies that $X(z,.)$ is admissible as well.

Conversely, let $X$ be a kernel such that $X(z,{\cdot})$ and 
$\ol{X}(\cdot,z)$ are admissible for all $z\in Z$. Then for fixed 
$z_\ell$,
\[
\sum_k c_k|z_k'\>=0 \implies \sum_k c_kX(z_\ell,z_k')=0,
\]
and for fixed $z_k'$,
\[
\sum_\ell c_\ell|z_\ell\>=0 \implies 
\sum_\ell\ol{c_\ell}X(z_\ell,z_k')=0. 
\]
Therefore, for given vectors $\phi=\D\sum_k c_k'|z_k'\>$ and
$\psi=\D\sum_\ell c_\ell|z_\ell\>\in\Qz(Z)$, the double sum 
\[
(\psi,\phi)_X:=\sum_\ell\sum_k\ol{c_\ell}{c_k}X(z_\ell,z_k)
\]
is independent of the representation of $\phi$ and $\psi$, hence 
defines a sesquilinear form. Thus 
\[
\psi\to\psi^*\X\phi:=(\psi,\phi)_X,\Forall\psi\in\Qz(Z)
\]
defines an antilinear functional $\X\phi:\Qz(Z)\to\Cz$. Thus
$\X\phi\in\Qz(Z)^\*$. Clearly, $\phi\to\X\phi$ defines a linear map
$\X:\Qz(Z)\to\Qz(Z)^\*$. Therefore $\X\in\Linx\Qz(Z)$. It is easy to 
check that $\<z|\X|z'\>=X(z,z')$ for all $z,z'\in Z$.
Thus $\sh \X=X$. Finally, it can be readily checked that $\X$ is
the unique operator which satisfies $X=\sh \X$. 
\epf

\section{Coherent maps and their quantization}\label{s.cohMaps}

This section discusses symmetries of a coherent space, one of the most
important concepts for studying and using coherent spaces. Indeed,
most of the applications of coherent spaces in quantum mechanics and
quantum field theory rely on the presence of a large symmetry group.
The main reason is that -- as we show in Theorem \ref{t.Gamma} below --
there is a quantization map that furnishes a representation of the
semigroup of coherent maps on the quantum space, and thus provides easy
access to a class of very well-behaved linear operators on the quantum
space.

Let $Z,Z'$ be coherent spaces. Recall from \sca{Neumaier} 
\cite[Section 5.3]{Neu.CQP} that a \bfi{morphism} from $Z$ to $Z'$ is a 
map $\rho:Z\to Z'$ such that
\lbeq{mor.coh}
K'(\rho z,\rho w)=K(z,w) \for  z,w\in Z;
\eeq
if $Z'=Z$, $\rho$ is called an \bfi{endomorphism}.
Two coherent spaces $Z$ and $Z'$ are called \bfi{isomorphic} if there is
a bijective morphism $\rho:Z\to Z'$. In this case we write $Z\cong Z'$
and we call the map $\rho:Z\to Z'$ an \bfi{isomorphism} of the coherent
spaces. Clearly, $\rho^{-1}:Z'\to Z$ is then also an isomorphism.

In the spirit of category theory one should define the symmetries of a
coherent space $Z$ in terms of its \bfi{automorphisms}, i.e.,
isomorphisms from $Z$ to itself. Remarkably, however, coherent spaces
allow a significantly more general concept of symmetry, based on the
notion of a coherent map.

\subsection{Coherent maps}\label{ss.cohMaps}

Let $Z$ and $Z'$ be coherent spaces with coherent products $K$ and $K'$,
respectively.
A map $A:Z'\to Z$ is called \bfi{coherent} if there is an
\bfi{adjoint map} $A^*:Z\to Z'$ such that
\lbeq{e.cohadj}
K(z,Az')=K'(A^*z,z') \for z\in Z,~z'\in Z'
\eeq
If $Z'$ is nondegenerate, the adjoint is unique, but not in general.
A coherent map $A:Z'\to Z$ is called an \bfi{isometry} if it has an
adjoint satisfying $A^*A=1$.
A \bfi{coherent map} on $Z$ is a coherent map from $Z$ to itself.

A \bfi{symmetry} of $Z$ is an invertible coherent map on $Z$ with an
invertible adjoint.
We call a coherent map $A$ \bfi{unitary} if it is invertible and
$A^*=A^{-1}$. Thus unitary coherent maps are isometries.

\begin{expl}
An \bfi{orbit} of a group $\Gz$ acting on a set $S$ is a set consisting
of all images $Ax$ ($A\in\Gz$) of a single vector. The group is
\bfi{transitive} on $S$ if $S$ is an orbit. The orbits of groups
of linear self-mappings of a Euclidean space give coherent spaces with
predefined transitive symmetry groups. Indeed, in the coherent space
formed by an arbitrary subset $Z$ of a Euclidean space $U$ with coherent
product $K(z,z'):=z^*z'$, all linear operators $A:U\to U^\*$ such that 
$A$ and $A^*$ map $Z$ into itself are coherent maps, and all such
operators mapping $Z$ bijectively onto itself are symmetries. This is 
the reason why coherent spaces are important in the theory of group 
representations.

For example, the symmetric group $\Sym(5)$ acts as a group of
Euclidean isometries on the 12 points of the icosahedron in $\Rz^3$.
The coherent space consisting of these 12 points with the induced
coherent product therefore has $\Sym(5)$ as a group of unitary
symmetries. The skeleton of the icosahedron is a distance-regular graph,
here a double cover of the complete graph on six vertices. Many other
interesting examples of finite coherent spaces are related to Euclidean 
representations of distance regular graphs (\sca{Brouwer} et al. 
\cite{BroCN}) and other highly symmetric combinatorial objects.
\end{expl}

\begin{prop}\label{p.uSym}~\\
(i) Every unitary coherent map is a symmetry.

(ii) An invertible map $A:Z\to Z$ is a unitary coherent map iff
\[
K(Az,Az')=K(z,z') \Forall z,z'\in Z.
\]
\end{prop}

\bepf
(i) $A^{-1}$ exists and is coherent by the preceding since $A^{-1}=A^*$.

(ii) Replace $z$ in \gzit{e.cohadj} by $Az$.
\epf

\begin{prop}~\\
(i) Every morphism $A$ with right inverse $A'$ is coherent, with
adjoint $A^*=A'$.

(ii) Every isometry is a morphism.

(iii) A map $A:Z\to Z$ is an automorphism of $Z$
iff it is a unitary coherent map.
\end{prop}

\bepf
(i) Put $A^*:=A'$. Then $AA^*=1$, and we have
$K(z,Az')=K(AA^*z,Az')=K(A^*z,z')$, proving the claim.

(ii) Let $A:Z\to Z'$ be an isometry. Then, for $z,z'\in Z$,
\[
K'(Az,Az')=K(z,z').
\]
(iii) Let $A:Z\to Z$ be an automorphism of $Z$.
Since $A$ is a morphism and invertible, for $z,z'\in Z$, we get
\[
K(z,Az')=K(AA^{-1}z,Az')=K(A^{-1}z,z').
\]
This implies that $A$ is coherent with $A^*:=A^{-1}$ with $A^*A=AA^*=1$.
Hence $A$ is unitary.
Conversely, assume that $A:Z\to Z$ is a unitary coherent map.
Then Proposition \ref{p.uSym}(ii) implies that $A$ is a morphism.
Since $A$ is bijective, it is an automorphism of $Z$ as well.
\epf

\begin{prop}\label{obs.0}
Let $Z$ be a coherent space and $A:Z\to Z$ be a coherent map. Then
for $z,z'\in Z$,
\lbeq{e.KAz}
K(Az,z')=K(z,A^*z'),
\eeq
\lbeq{e.Az}
\<z|Az'\>=\<A^*z|z'\>,~~~\<Az|z'\>=\<z|A^*z'\>.
\eeq
\end{prop}

\bepf
For $z,z'\in Z$, \gzit{e.Kherm} implies
\[
\<Az|z'\>=K(Az,z')=\ol{K(z',Az)}=\ol{K(A^*z',z)}=K(z,A^*z')
=\<z|A^*z'\>.
\]
This proves both \gzit{e.KAz} and the second half of \gzit{e.Az}.
The first half of \gzit{e.Az} follows directly from \gzit{e.cohadj}.
\epf

\begin{thm}\label{t.cohSemi}
Let $Z$ be a coherent space. Then the set $\coh Z$  consisting of all
coherent maps is a semigroup with identity. Moreover:

(i) Any adjoint $A^*$ of $A\in\coh Z$ is coherent.

(ii) For any invertible coherent map $A:Z\to Z$ with an invertible
adjoint, the inverse $A^{-1}$ is coherent.
\end{thm}

\bepf
The identity map $I:Z\to Z$ is trivially
coherent. Let $A,B\in\coh Z$. Then, for $z,z'\in Z$,
\[
K(z,ABz')=K(A^*z,Bz')=K(B^*A^*z,z'),
\]
which implies that $AB$ is coherent with adjoint $(AB)^*=B^*A^*$.

(i) Using Proposition \ref{obs.0}, we can write
\[
K(z,A^*z')=K(Az,z'),
\]
which implies that $A^*$ is coherent with $A^{**}=A$.

(ii)  Let $A:Z\to Z$ be a coherent map with an adjoint $A^*$ such that
$A$ and $A^*$ are invertible with the inverses $A^{-1}$ and
$(A^*)^{-1}$. Then, for $z,z'\in Z$,
\[
 K(A^{-1}z,z')=K(A^{-1}z,A^*(A^*)^{-1}z')=K(AA^{-1}z,(A^*)^{-1}z')
=K(z,(A^*)^{-1}z'),
\]
which implies that $A^{-1}$ is coherent with $(A^{-1})^*=(A^*)^{-1}$.
\epf

\begin{cor}\label{c.*semi}
Let $Z$ be a nondegenerate coherent space. Then $\coh Z$ is a
$*$-semigroup with identity, i.e.,
\[
1^*=1,~~~A^{**}=A,~~~(AB)^*=B^*A^* \for A,B\in\coh Z.
\]
Moreover, the set $\Sym(Z)$ of all invertible coherent maps with
invertible adjoint is a $*$-group, and
\[
A^{-*}:=(A^{-1})^*=(A^*)^{-1} \for A\in\Sym(Z).
\]
\end{cor}

\bepf
If $Z$ is nondegenerate then the adjoint is unique. Therefore the
claim follows from the preceding result.
\epf

As a consequence, the vector space spanned by all coherent maps of a 
nondegenerate coherent space has a natural $*$-algebra structure in 
the sense of \sca{Powers} \cite{Pow}.

For any coherent space $Z$, $[Z]$ denotes the nondegenerate coherent
space defined in \cite[Proposition 5.8.7]{Neu.CQP}, with the same 
quantum space as $Z$.

\begin{prop}
Let $Z$ be a coherent space. Then
\[
\coh [Z]=\{[A] \mid A\in\coh Z\}.
\]
\end{prop}
\bepf
Let $A\in\coh Z$. Then $[A]\in\coh [Z]$ by
\cite[Theorem 5.8.9]{Neu.CQP}.
Thus $\{[A]:A\in\coh Z\}\subseteq\coh [Z]$.
Let $\imath:[Z]\to Z$ be a choice function, that is a function which
satisfies $[\imath[z]]=[z]$ for all $z\in Z$. For any coherent map
$\mathcal{A}:[Z]\to[Z]$, define $A:Z\to Z$ by
$z\to Az:=\imath(\mathcal{A}[z])$. Then $A:Z\to Z$ is a well-defined
map.  Thus, for $z,z'\in Z$,
\[
\bary{lll}
K(Az,z')
&=&K(\imath(\mathcal{A}[z]),z')=K([\imath(\mathcal{A}[z])],[z'])
=K(\mathcal{A}[z],[z'])\\
&=&K([z],\mathcal{A}^*[z'])=K([z],[\imath(\mathcal{A}^*[z'])])
=K(z,\imath(\mathcal{A}^*[z'])).
\eary
\]
This implies that $A$ is a coherent map with an adjoint $A^*:Z\to Z$
given by $A^*z=\imath(\mathcal{A}^*[z'])$. For $z,z'\in Z$, we have
\[
K([A][z],[z'])=K([Az],[z'])=K([\imath(\mathcal{A}[z])],[z'])
=K(\mathcal{A}[z],[z']),
\]
implying that $[A]=\mathcal{A}$.
\epf

\begin{thm}\label{t.PA}
Let $PZ$ be the projective extension of degree 1 of the coherent space
$Z$.

(i) Let $A:Z\to Z$ be a map with the property
\lbeq{e.cohMult}
K(z, Az')v(z') = \ol{w(z)}K(A^*z, z') \for z,z'\in Z,
\eeq
for suitable $v,w:Z\to\Cz$ and $A^*:Z\to Z$. Then
\[
[\alpha,A](\lambda,z):=(\alpha v(z)\lambda,Az),~~~
[\alpha,A]^*(\lambda,z):=(\ol\alpha w(z)\lambda,A^*z)
\]
define a coherent map $[\alpha,A]$ of $PZ$ and its adjoint
$[\alpha,A]^*$.

(ii) For every coherent map $A:Z\to Z$ and every $\alpha\in\Cz$, the map
$[\alpha,A]:PZ\to PZ$ defined by
\[
[\alpha,A](\lambda,z)
:=(\alpha\lambda,Az)\Forall (\lambda,z)\in PZ,
\]
is coherent.
\end{thm}

\bepf
Let $(\lambda,z),(\lambda',z')\in PZ$. Then
\[
K_{\rm pe}((\lambda,z),(\lambda',z'))=\ol\lambda K(z,z')\lambda'.
\]
Therefore,
\[
\bary{lll}
K_{\rm pe}((\lambda,z),[\alpha,A](\lambda',z'))
&=&K_{\rm pe}((\lambda,z),(\alpha v(z')\lambda',Az'))

=\ol{\lambda}K(z,Az')\alpha v(z')\lambda'\\
&=&\ol{\lambda\ol\alpha}K(z,Az') v(z')\lambda'
=\ol{\lambda\ol\alpha w(z)}K(A^*z,z')\lambda'\\
&=&K_{\rm pe}((\lambda\ol\alpha w(z),A^*z),(\lambda',z'))
=K_{\rm pe}([\alpha,A]^*(\lambda,z),(\lambda',z')).
\eary
\]
This proves (i), and (ii) is the special case of (i) where $v$ and $w$
are identically 1.
\epf

Something similar can be shown for projective extensions of any integral
degree $e\ne 0$. Condition \gzit{e.cohMult} appears first in a paper 
by \sca{Bertram \& Hilgert} \cite{BerH} on reproducing kernels invariant
under an involutive semigroup.

\subsection{Some examples}\label{ss.Szego}

Coherent spaces with the same quantum space can have very different 
symmetry groups. Typically, the largest groups are associated with 
projective coherent spaces. We illustrate this here with two simple 
coherent spaces. Another large class of examples of this situation
is treated extensively in Section \ref{s.Klauder}.

\begin{expl}\label{ex.Szego}
(\sca{Szeg\"o} \cite{Sze}, 1911)
The \bfi{Szeg\"o space} is the coherent space $Z$ defined on the
open unit disk in $\Cz$,
\[
D(0,1):=\{z\in\Cz\mid |z|<1\},
\]
by the coherent product
\[
K(z,z'):=(1-\ol zz')^{-1}.
\]
the inverse is defined since $|\ol zz'|<1$. A corresponding quantum
space is the \bfi{Hardy space} of power series
\[
f(x)=\sum_{\ell=0}^\infty f_\ell x^\ell
\]
such that
\[
\|f\|:=\sqrt{\sum |f_\ell|^2}<\infty,
\]
describing analytic functions on $Z$ that are square integrable over the
positively oriented boundary $\partial Z$ of $Z$, with inner product
\[
f^*g:=\sum \ol f_\ell g_\ell
=\int_0^{2\pi} d\phi \ol{f(e^{i\phi})}g(e^{i\phi})
=\int_{\partial Z} |dz| \ol{f(z)}g(z).
\]
The associated coherent states are the functions
\[
k_z(x)=(1-zx)^{-1}
\]
with $(k_z)_\ell=z^\ell$, since
\[
k_z^*k_{z'}=\sum \ol z^\ell(z')^\ell=\frac{1}{1-\ol zz'}=K(z,z').
\]
Clearly, the scalar multiplication maps $z\to\lambda z$ for 
$\lambda\in D(0,1)$ are coherent, with the complex conjugate as adjoint.
There are no other coherent maps with adjoints. Indeed, suppose 
$A:Z\to Z$ is a coherent map with adjoint $A^*$. Then for every 
$z,z'\in D(0,1)$,
\[
\ol{Az}z'=1-K(Az,z')^{-1}=1-K(z,A^*z')^{-1}=\ol{z}A^*z'. 
\]
Thus $Az=\lambda z$ for $\lambda:=\ol{A^*z'/z'}$ with any fixed nonzero 
$z'$. 
\end{expl}

\begin{expl}\label{ex.Moebius}
The \bfi{M\"obius space} $Z:=\{z\in\Cz^2 \mid |z_1|>|z_2|\}$ is a
coherent space with coherent product
\[
K(z,z'):=(\ol z_1z_1'-\ol z_2z_2')^{-1}
\]
with the same quantum spaces as the Szeg\"o space. Indeed, the functions
\lbeq{e.Mcoh}
f_z(x)=(z_1-z_2x)^{-1}
\eeq
from the Szeg\"o space from Example \ref{ex.Szego} are associated
\bfi{M\"obius coherent states}.
The M\"obius space is a projective coherent space of degree $-1$;
indeed, with the scalar multiplication induced from $\Cz^2$, we have
\[
K(z,\lambda z')=(\ol z_1\lambda z_1'-\ol z_2\lambda z_2')^{-1}
=\lambda^{-1}(\ol z_1z_1'-\ol z_2z_2')^{-1}=\lambda^{-1} K(z,z')
\]
for all $z,z'\in Z$. It is now easy to see that the projective
completion of the Szeg\"o space for this degree is isomorphic to the
M\"obius space.

Unlike the Szeg\"o space, the M\"obius space has a large symmetry group.
Indeed, for any $A\in\Cz^{2\times 2}$ we have
\lbeq{e.Az12}
|(Az)_1|^2-|(Az)_2|^2=\alpha|z_1|^2+2\re(\beta\ol z_1z_2)-\gamma|z_2|^2
\eeq
with
\[
\alpha:=|A_{11}|^2-|A_{21}|^2,~~~
\beta:=\ol A_{11}A_{12}-\ol A_{21}A_{22},~~~
\gamma:=|A_{22}|^2-|A_{12}|^2,
\]
If $z\in Z$ and
\lbeq{e.Aineq}
\alpha>0,~~~|\beta|\le \alpha,~~~\gamma\le \alpha-2|\beta|
\eeq
we write $\beta=|\beta|\delta$ with $|\delta|=1$ and obtain from
\gzit{e.Az12}
\[
\bary{lll}
|(Az)_1|^2-|(Az)_2|^2
&=&\alpha|z_1|^2+2|\beta|\re(\delta\ol z_1z_2)-\gamma|z_2|^2\\
&\ge&
  \alpha|z_1|^2+2|\beta|\re(\delta\ol z_1z_2)+(2|\beta|-\alpha)|z_2|^2\\
&=& |\beta|\,|z_1+\delta z_2|^2+(\alpha-|\beta|)(|z_1|^2-|z_2|^2) \ge 0.
\eary
\]
Equality in the last step is possible only if $|\beta|=\alpha>0$ and
$z_1+\delta z_2=0$, contradicting $|z_1|>|z_2|$. Hence $Az\in Z$.
Thus $A$ maps $Z$ into itself whenever \gzit{e.Aineq} holds.
Now
\[
\bary{lll}
K(z,Az')
&=&\Big(\ol z_1(A_{11}z_1'+A_{12}z_2')
        -\ol z_2(A_{21}z_1'+A_{22}z_2')\Big)^{-1}
\\
&=&\Big((A_{11}\ol z_1-A_{21}\ol z_2)z_1'
        -(-A_{12}\ol z_1+A_{22}\ol z_2)z_2'\Big)^{-1}
=K(A^\sigma z,z'),
\eary
\]
where
\[
A^\sigma=\pmatrix{ \ol A_{11} & -\ol A_{21} \cr
                  -\ol A_{12} &  \ol A_{22}}.
\]
Thus every linear mapping $A\in\Cz^{2\times 2}$ satisfying
\gzit{e.Aineq} is coherent, with adjoint $A^*$ given by $A^\sigma$
rather than by the standard matrix adjoint. These mappings form a
semigroup, a homogeneous version of an Olshanski semigroup of
compressions (\sca{Olshanskii} \cite{Ols}). By \gzit{e.Az12}, 
$A$ preserves the Hermitian form $|z_1|^2-|z_2|^2$ up to a positive 
scaling factor iff $\beta=0$ and $\gamma=\alpha>0$. This implies 
\gzit{e.Aineq} and is equivalent with
\lbeq{e.Agroup}
\ol A_{11}A_{12}=\ol A_{21}A_{22},~~~|A_{22}|=|A_{11}|>|A_{21}|.
\eeq
Indeed, the first equation and the inequality follow from $\beta=0$ and 
$\alpha>0$. From the first equation, 
\[
\alpha=\gamma=|A_{22}|^2-|A_{12}|^2
=|A_{22}|^2-\Big(|A_{21}||A_{22}|/|A_{11}|\Big)^2
=\alpha\Big(|A_{22}|/|A_{11}|\Big)^2,
\]
giving $|A_{22}|=|A_{11}|$. Thus the group $GU(1,1)$ of all matrices
satisfying \gzit{e.Agroup} is a group of symmetries of $Z$. This fact 
is relevant for applications to quantum systems with a dynamical 
symmetry group $SU(1,1)$ or the closely related groups $SO(2,1)$, 
$SL(2,\Rz)$, which are subgroups of $GU(1,1)$.
\end{expl}

\subsection{Quantization of coherent maps}

\begin{thm}\label{t.Gamma}
Let $Z$ be a coherent space, $\Qz(Z)$ a  quantum space
of $Z$, and let $A$ be a coherent map on $Z$.

(i) There is a unique linear map $\Gamma(A)\in\Lin\Qz(Z)$ such that
\lbeq{e.move}
\Gamma(A)|z\>=|Az\>\Forall z\in Z.
\eeq
(ii) For any adjoint map $A^*$ of $A$,
\lbeq{e.zGamma}
\<z|\Gamma(A)=\<A^*z| \Forall z\in Z,
\eeq
\lbeq{e.GammaStar}
\Gamma(A)^*|_{\Qz(Z)}=\Gamma(A^*).
\eeq
(iii) The definition $\Gamma(A):=\Gamma(A^*)^*$ extends $\Gamma(A)$ to 
a linear map $\Gamma(A)\in \Lin\Qz^\*(Z)$.
\end{thm}

We call $\Gamma(A)$ and its extension the \bfi{quantization}
of $A$ and $\Gamma$ the \bfi{quantization map}.
In the special case (discussed in Section \ref{s.Klauder} below) where 
$Z$ is a Klauder space, the quantum space $\Qz(Z)$ is a dense subspace 
of a bosonic Fock space and the quantization map is the restriction of 
the \bfi{second quantization} map of \sca{Derezi\'nski \& G\'erard} 
\cite{DerG} to $\Qz(Z)$. This terminology goes back to \sca{Fock} 
\cite{Fock}.

\bepf
(i) Let $A:Z\to Z$ be a coherent map and $S:Z\times Z\to\Cz$
be the kernel given by
\[
S(z,z'):=K(z,Az') \Forall z,z'\in Z.
\]
We first show that for all $z\in Z$, $S(\cdot,z)$ and $\ol{S}(z,\cdot)$ 
are admissible functions. Suppose that $\sum_\ell c_\ell|z_\ell\>=0$. 
Then
\[
\sum\ol{c_\ell}S(z_\ell,z')=\sum\ol{c_\ell}K(z_\ell,Az') 
=\Big\<\sum c_\ell z_\ell\Big|Az'\Big\>=0,
\]
proving that $S(\cdot,z)$ is admissible. Similarly,
\[
\bary{lll}
\sum\ol{c_\ell}\ol S(z,z_\ell)
&=&\D\sum\ol{c_\ell}\ol{K(z,Az_\ell)}
=\sum\ol{c_\ell}K(Az_\ell,z) \\
&=&\D\sum\ol{c_\ell}K(z_\ell,A^*z) 
=\Big\<\sum c_\ell z_\ell\Big|A^*z\Big\>=0,
\eary
\]
proving that $\ol{S}(z,\cdot)$  is admissible. 
By Theorem \ref{t.opExist}, there is a 
unique linear operator $\Gamma(A):\Qz(Z)\to\Qz(Z)^\*$ satisfying
\lbeq{gam.main}
S(z,z')=\<z|\Gamma(A)|z'\>\Forall z,z'\in Z,
\eeq
and it is automatically continuous.
To prove the theorem we need to show that the images are
actually in $\Qz(Z)$. Using \gzit{e.move}, we have
\lbeq{qz.to.qz}
\<z|Az'\>=K(z,Az')=S(z,z')=\<z|\Gamma(A)|z'\>\Forall z,z'\in Z.
\eeq
which implies that $\Gamma(A)|z'\>=|Az'\>$ for all $z'\in Z$.
We conclude that $\Gamma(A)$ maps $\Qz(Z)$ already into the smaller
space $\Qz(Z)$. Hence $\Gamma(A)\in\Lin\Qz(Z)$.

(ii) Let $z\in Z$ and $\phi=\D\sum c_k|z_k\>\in\Qz(Z)$.
Then \gzit{e.zGamma} follows from
\[
\bary{lll}
 \Big(\Gamma(A)^*\<z|\Big)(\phi)
&=&\D\<z|\Gamma(A)\phi
 =\<z|\Gamma(A)\sum c_k|z_k\>
=\<z|\sum c_k\Gamma(A)|z_k\>\\[4mm]
&=&\D\<z|\sum c_k|Az_k\>
=\sum c_k\<z|Az_k\>
=\sum c_k\<A^*z|z_k\>\\[4mm]
&=&\D\<A^*z|\sum c_k|z_k\>=\<A^*z|\phi,
\eary
\]
By Theorem \ref{t.cohSemi}(i), the map $A^*$ is coherent as well.
Thus we have
\[
\bary{lll}
\<z|\Gamma(A)^*|z'\>
&=&\ol{\<z'|\Gamma(A)|z\>}=\ol{\<z'|Az\>}=\ol{K(z',Az)}\\
&=&K(Az,z')=K(z,A^*z')
=\<z|A^*z'\>=\<z|\Gamma(A^*)|z'\>,
\eary
\]
which implies that the restriction of $\Gamma(A)^*$ to $\Qz(Z)$ is
precisely $\Gamma(A^*)$, as claimed.

(iii) is a simple consequence of (i) and (ii).
\epf

We now show that the quantization map $\Gamma$ furnishes a
representation of the semigroup of coherent maps on $Z$ in the quantum
space of $Z$.

\begin{thm}\label{t.Gamma2}
The quantization map $\Gamma$ has the following properties.

(i) The identity map $1$ on $Z$ is coherent, and $\Gamma(1)=1$.

(ii) For any two coherent maps $A,B$ on $Z$,
\[
\Gamma(AB)=\Gamma(A)\Gamma(B).
\]
(iii) For any invertible coherent map $A:Z\to Z$ with an invertible
adjoint, $\Gamma(A)$ is invertible with inverse
\[
\Gamma(A)^{-1}=\Gamma(A^{-1}).
\]
(iv) For a coherent map $A:Z\to Z$, $A$ is unitary iff $\Gamma(A)$ is
unitary.
\end{thm}

\bepf
(i) is straightforward.

(ii) Let $A,B$ be coherent maps and $z,z'\in Z$.
Then we have
\[
 \<z|\Gamma(AB)|z'\>=K(z,ABz')=\<z|\Gamma(A)|Bz'\>
=\<z|\Gamma(A)\Gamma(B)|z'\>,
\]
which implies that $\Gamma(AB)=\Gamma(A)\Gamma(B)$.

(iii) follows from $\Gamma(1)=1$ and the fact that $AA^{-1}=A^{-1}A=1$.
Indeed, using Theorem \ref{t.cohSemi}(ii), $A^{-1}$ is coherent and we
get
\[
\Gamma(A)\Gamma(A^{-1})=\Gamma(AA^{-1})=\Gamma(1)=\Gamma(A^{-1}A)=
\Gamma(A^{-1})\Gamma(A),
\]
which implies that $\Gamma(A)$ is invertible with
$\Gamma(A)^{-1}=\Gamma(A^{-1})$.

(iv) Let $A$ be a coherent map. Also, suppose that $A$ is unitary as
well. Then, $A$ is invertible with the inverse $A^{-1}=A^*$.
Thus, $A$ and $A^*$ are invertible. Then, we get
\[
 \Gamma(A)\Gamma(A)^*=\Gamma(A)\Gamma(A^*)=\Gamma(AA^*)=\Gamma(1)=1,
\]
and also
\[
\Gamma(A)^*\Gamma(A)=\Gamma(A^*)\Gamma(A)=\Gamma(A^*A)=\Gamma(1)=1.
\]
Hence, we deduce that $\Gamma(A)$ is a unitary linear operator.
Conversely, assume that $\Gamma(A)$ is a unitary linear operator.
Then we get $AA^*=1$ and also $A^*A=1$, which means that $A$ is unitary.
\epf

The quantization map is important as it reduces many computations
with coherent operators in the quantum space of $Z$ to computations in
the coherent space $Z$ itself. By Theorem \ref{t.Gamma2}, large
semigroups of coherent maps $A$ produce large semigroups of coherent
operators $\Gamma(A)$, which may make complex calculations much more
tractable.

\subsection{Geometric quantization and quantum field theory}
\label{s.geomQ}

In this subsection we discuss informally connections between coherent 
spaces and topics from geometric quantization and quantum field theory. 
Later, we give full details for one particular case, the case of Klauder
spaces and their connection to bosonic Fock spaces. Details 
regarding the other issues will be discussed elsewhere 
(\sca{Neumaier} \cite{Neu.cohFock}).

\bigskip

{\bf Line bundles and central extensions.}
Example \ref{ex.Moebius} generalizes to central extensions of other 
semisimple Lie groups and associated line bundles over symmetric spaces.
This follows from the material on the corresponding coherent states 
discussed in detail in \sca{Perelomov} \cite{Per} from a group 
theoretic point of view, and in \sca{Zhang} et al. \cite{ZhaFG} in 
terms of applications to quantum mechanics. Coherent spaces with the 
structure of a vector bundle also accommodate the vector coherent 
states of \sca{Rowe} et al. \cite{RowRG} and \sca{Bertram \& Hilgert} 
\cite{BerH}. Other related material is in the books by 
\sca{Faraut \& Kor\'anyi} \cite{FarK}, \sca{Neeb} \cite{Nee}, and 
\sca{Neretin} \cite{Ner}.

In the applications, a group $\Gz$ of quantum symmetries is typically
first defined classically on a symmetric space. In a quantization step,
it is then represented by a unitary representation on a Hilbert space.
Typically, unitary representations  of the symmetry groups of symmetric
spaces are only projective representations, defined in terms of a
family of multipliers satisfying a cocycle condition. This central 
extension (defined through the respective cocycle) is represented 
linearly in the Hilbert space defined through the geomentric 
quantization procedure. 

\bigskip

{\bf K\"ahler potentials.}
Therefore, in geometric quantization (\sca{Woodhouse} \cite{Woo}), the 
symmetric space (typically a K\"ahler manifold) needs to be extended to 
a line bundle on which a central extension of the group acts 
classically. 
This line bundle can be turned invarious ways into coherent spaces by 
defining the coherent product as the exponential of suitable K\"ahler 
potentials. For the symmetric spaces associated to finite-dimensional 
semisimple Lie groups, the corresponding unitary representations and 
their K\"ahler potentials are constructed in papers by 
\sca{Bar Moshe \& Marinov} \cite{BarM1,BarM2}. 

In the coherent space setting, the coherent product defined on an orbit
$Z$ of $\Gz$ on the symmetric space $Z$ via the coherent states
available from geometric quantization leads in these cases to a
coherent space. However, on this space, most elements of $\Gz$ are not
represented coherently since they only satisfy a relation
\gzit{e.cohMult} with multipliers that are not constant.
Theorem \ref{t.PA} shows that the projective extension $PZ$ of degree 
$1$ represents the central extension coherently. This shows that
projective coherent spaces are the natural starting point for coherent
quantization since they represent all classically visible symmetries
in a coherent way. The projective property is therefore typically
needed whenever one has a quantum system given in terms of a coherent
space and wants to describe all symmetries of the quantum system through
coherent maps.

\bigskip

{\bf Bosonic Fock spaces.}
A very important class of Hilbert spaces is the family of bosonic Fock 
spaces (\sca{Fock} \cite{Fock}). They are indispensable in quantum field
theory (\sca{Baez} et al. \cite{BaeSZ}, \sca{Derezi\'nski \& G\'erard} 
\cite{DerG}, \sca{Glimm \& Jaffe} \cite{GliJ})
and the theory of Hida distributions in the white noise calculus
for classical stochastic processes (\sca{Hida \& Si} \cite{HidS},
\sca{Hida \& Streit} \cite{HidSt}, \sca{Obata} \cite{Oba}).
In Section \ref{s.Klauder} we discuss in some detail Klauder spaces, a 
class of coherent spaces whose completed quantum spaces of Klauder 
coherent spaces are shown in Section \ref{ss.bFock} to be the bosonic 
Fock spaces. We identify operators on these quantum spaces 
corresponding to creation and annihilation operators in Fock space, 
and prove their basic properties.
In particular, we prove the Weyl relations, the canonical commutation
relations, without the need to know a particular realization of the
quantum space. We also show that the abstract normal ordering,
introduced in Subsection \ref{ss.normal} below, reduces for Klauder 
spaces to that familiar from traditional second quantization. 
Klauder spaces have a large semigroup of coherent maps; their 
quantization defines normally ordered exponentials of certain 
inhomogenoeus quadratics in $a^*$ and $a$ corresponding to a particular 
class of Bogoliubov transformations in quantum field theory.

\bigskip

{\bf Squeezed states and metaplectic groups.}
Bosonic Fock spaces are also completed quantum spaces of a class of 
coherent spaces containing the labels for all squeezed states (cf.
\sca{Zhang} et al. \cite{ZhaFG} for the case of finitely many modes, 
and \sca{V\'arilly \& Gracia-Bond\'ia} et al. \cite{VarG} for 
the case of infinitely many modes). Geometrically, these coherent 
spaces are complex line bundles over symmetric spaces carrying a 
unitary representation of a metaplectic group. Therefore the 
metaplectic groups are realized by coherent maps on these coherent 
spaces. Their quantization leads to normally ordered exponentials of 
all sufficiently regular homogenoeus quadratics in $a^*$ and $a$. 
Even larger coherent spaces contain coherent maps realizing the 
semidirect product of the metaplectic group with certain Heisenberg 
groups. Their quantization leads to normally ordered exponentials of 
all sufficiently regular inhomogenoeus quadratics in $a^*$ and $a$. 

\bigskip

{\bf Fermionic Fock spaces and spin groups.}
Statements analogous to those for metaplecic groups and bosonic Fock 
spaces can be proved for spin groups and fermionic Fock spaces.
Now the coherent spaces, which we call \bfi{Hua spaces} for reasons 
discussed in a moment, are complex line bundles over symmetric spaces 
carrying a unitary representation of a spin group. Therefore the 
spin groups are realized by coherent maps on Hua spaces. 
Their quantization again leads to normally ordered exponentials of 
all sufficiently regular homogenoeus quadratics in $a^*$ and $a$. 

Traditionally, the fermionic Fock space construction is based on
a distinguished vacuum state -- the 0-particle state. However, in
quantum field theory, many vectors in a fermionic Fock space may serve 
as a potential \bfi{vacuum state}; the vacuum of a Fock space depends 
on the Hamiltonian under consideration. In time-dependent systems, the 
Hamiltonian and hence the corresponding vacuum state changes with time. 
Thus the structure of interest in quantum field theory is not the 
Fock space itself but a mathematically precise version of ''what is 
left from Fock space when no vacuum state is distinguished''. This Fock 
space stripped of a special vacuum state is, as a vector space, the 
same object as the Fock space itself, but each choice of a vacuum 
defines a different associated multi-particle structure, anticommuting 
algebra, and exterior algebra, related to each other by Bogoliubov 
transformations.

This becomes transparently encoded in terms of the associated Hua
spaces. Seen abstractly, these are line bundles of symmetric spaces 
without any distinguished origin. This like using an affine space where 
all points are on equal footing, as compared to a vector space that has 
a distinguished origin.
Just as in the case of affine spaces, its coordinatization requires the 
choice of an origin. Its stabilizer defines (in the simplest case of 
Fock spaces with only finitely many modes) a particular group $U(n)$ 
that gives rise to an identification of the symmetric space with a
homogeneous space $SO(2n)/U(n)$ in the coordinates used by \sca{Zhang} 
et al.. However, just as in the case of affine spaces, this choice of
origin is immaterial due to the presence of coherent maps that move the 
origin to an arbitrary point. In particular, the associated quantum 
space has no distinguished state vector -- only the whole class of 
coherent states is distinguished, and any of them may serve as a 
potential vacuum state. 

In the case of charged matter in a time-dependent external 
electromagnetic field (the only case whose quantization is truly 
understood,\sca{Gracia-Bond\'ia \& V\'arilly} et al. \cite{GraV}), the 
single-particle dynamics is given by a linear Dirac equation in this 
Euclidean space with (in general) time-dependent Hamiltonian $H(t)$, 
called a \bfi{Dirac operator}. In the language of second quantization, 
this Euclidean space is called the 1-particle space, although its 
classical meaning is that of a space of half-densities for charged 
fermionic matter. As is well-known, a Dirac operator has a real 
spectrum unbounded in both directions.
At all times where this Dirac operator has no generalized eigenstate 
with zero energy, the invariant subspace spanned by the positive 
eigenfunctions of the Dirac operator is a maximal isotropic subspace 
of the classical symplectic form defined by the imaginary part of the
Hilbert inner product. Unless the Dirac operator happens to be time 
independent, this subspace changes with time, and defines the
coherent state specifying the correct vacuum state at each time $t$. 

Therefore the coherent rays $\Cz|z\>$ of a Hua space are in one-to one 
correspondence with the orthogonal projectors $P$ to a maximal isotropic
subspace of the Euclidean space. $1-P$ projects to a complementary 
maximal isotropic subspace.
A description of the symmetric space associated with $SO(2n)/U(n)$ 
in terms of such pairs of complementary maximal isotropic subspaces
was first given by \sca{Hua} \cite[Section II]{Hua}, who showed 
that they have the structure of a metric space with a distance function 
taking integer values only. Thus the points form an infinite bipartite 
graph, in modern terminology (see \sca{Brouwer} et al. \cite{BroCN}) 
called the \bfi{dual polar graph} $D_n(\Cz)$, and the $SO(2n)$ acts
as a distance transitive group of automorphism on each bipartite half.

A \bfi{Hua space} is a coherent space defined (in the finite mode case) 
on the canonically associated complex line bundle via a canonical 
K\"ahler structure. In the applications to quantum field theory, a 
nontrivial limit $n\to\infty$ must be taken, and one must consider 
infinite-dimensional Hua spaces. This limit $n\to\infty$ involves 
renormalization issues. With the current knowledge, one can do 
nonperturbative renormalization only in 2-dimensional quantum field 
theories, where normal ordering is all that is needed -- see 
\sca{Pressley \& Segal} \cite{PreS}.

\section{Homogeneous and separable maps}\label{s.hom}

In this section we look at self-mappings of coherent spaces satisfying
homogeneity or separability properties. These often give simple but
important coherent maps.

\subsection{Homogeneous maps and multipliers}\label{ss.hom}

Let $Z$ be a coherent space. We say that a function
$m:Z\to \Cz$ is a \bfi{multiplier} for the map $A:Z\to Z$ if
\lbeq{e.alphaA}
|z'\>=\lambda|z\> \implies m(z')|Az'\>=\lambda m(z)|Az\>,
\eeq
for all $\lambda\in\Cz$ and $z,z'\in Z$, equivalently if
\[
K(w,z')=\lambda K(w,z) ~\forall w\in Z \implies
m(z')K(w,Az')=\lambda m(z)K(w,Az)~\forall w\in Z.
\]
A function $m:Z\to\Cz$ is called \bfi{homogeneous} if
\lbeq{e.lambdam}
|z'\>=\lambda|z\>,~\lambda\ne 0 \implies m(z')=m(z);
\eeq
this is the case iff it is a multiplier for the identity map.

We call a map $A:Z\to Z$ \bfi{homogeneous} if
\lbeq{e.lambdaA}
|z'\>=\lambda|z\> \implies |Az'\>=\lambda |Az\>;
\eeq
this is the case iff $m=1$ is a multiplier for $A$. We write $\Hom Z$
for the set of all homogeneous maps $A:Z\to Z$.

\begin{thm}
Let $Z$ be a coherent space. Then,

(i) each coherent map is homogeneous.

(ii) the composition of any two homogeneous maps is homogeneous.
\end{thm}
\bepf
(i) Let $A$ be coherent map with an adjoint $A^*$. Suppose that
$z,z'\in Z$ and $\lambda\in\Cz^\*$ with $|z'\>=\lambda|z\>$.
Then, for $z''\in Z$, we get
\[
\<z''|Az'\>=\<A^*z''|z'\>=\lambda\<A^*z''|z\>=\lambda\<z''|Az\>.
\]
Thus $|Az'\>=\lambda|Az\>$. Therefore, $m=1$ is a multiplier for $A$
and hence $A$ is homogeneous.

(ii) Let $A,B\in\Hom(Z)$. Suppose that $z,z'\in Z$ and
$\lambda\in\Cz^\*$ with $|z'\>=\lambda|z\>$. Since $B$ is homogeneous,
we have $|Bz'\>=\lambda|Bz\>$.
Then applying homogeneity of $A$, we have $|ABz'\>=\lambda|ABz\>$.
Therefore, $m=1$ is a multiplier for $AB$ and hence $AB$ is homogeneous.
\epf

\begin{thm}
Let $Z$ be a projective coherent space. We then have
\[
K(z,\lambda z')=K(\ol{\lambda}z,z'),
\]
for all $z,z'\in Z$ and $\lambda\in\Cz^\*$. In particular, if $Z$ is a
nondegenerate and projective coherent space the
scalar multiplication map $\lambda:Z\to Z$ is coherent, with unique
adjoint $\lambda^*=\ol\lambda$.
\end{thm}

\bepf
Let $\lambda\in\Cz^\*$ be given. Then, for $z,z'\in Z$, we have
\[
K(\ol{\lambda}z,z')=\ol{K(z',\ol{\lambda}z)}
=\ol{\ol{\lambda}^eK(z',z)}=\lambda^e\ol{K(z',z)}
=\lambda^eK(z,z')=K(\lambda z,z').
\]
In particular, if $Z$ is nondegenerate then the multiplication map
$\lambda$ is coherent with the unique adjoint $\ol{\lambda}$.
\epf

\begin{prop}
Let $Z$ be a projective and non-degenerate coherent space. Then:

(i) $m:Z\to\Cz$ is a multiplier for $A:Z\to Z$ iff
\[
m(\mu z)|A\mu z\>=m(z)|\mu Az\> \Forall \mu\in\Cz^\*.
\]
(ii) A map $A:Z\to Z$ is homogeneous iff $A\mu=\mu A$ for all
$\mu\in\Cz^\*$.

(iii) A map $m:Z\to\Cz$ is homogeneous iff $m\mu=\mu m$ for all
$\mu\in\Cz^\*$.
\end{prop}

\bepf
In a projective coherent space of degree $e$, 
$|\lambda z\>=\lambda^e|z\>$. Nondegeneracy therefore implies that 
$|z'\>=\lambda|z\>$ iff $z'=\mu z$
for some choice of the $e$th root $\mu=\lambda^{1/e}$.
The definition of a multiplier now gives (i), and a straightforward
specialization gives (ii) and (iii).
\epf

\subsection{Factorizing maps}

Let $Z$ be a coherent space. We call a map $\alpha:Z\to Z$
\bfi{factorizing} if
there is a number $\chi(\alpha)\in\Cz$, called a
\bfi{factorization constant}, such that
\lbeq{chi.main}
K(z,\alpha z')=\chi(\alpha)K(z,z') \for z,z'\in Z.
\eeq

\begin{prop}
Let $Z$ be a coherent space and $\alpha:Z\to Z$ be a map. Then, $\alpha$
is factorizing iff there exists a complex constant $\lambda_\alpha$, 
such that for any quantum space $\Qz(Z)$ of $Z$ we have
\lbeq{equiv.sep}
|\alpha z\>=\lambda_\alpha|z\>\Forall z\in Z.
\eeq
In this case, $\chi(\alpha)=\lambda_\alpha$.
\end{prop}

\bepf
Let $\Qz(Z)$ be a quantum space of $Z$ and $z,z'\in Z$.
If $\alpha$ is factorizing with the factorization constant
$\chi(\alpha)$, then
\[
\<z'|\alpha z\>=K(z',\alpha z)=\chi(\alpha)K(z',z)
=\chi(\alpha)\<z'|z\>=\<z'|\Big(\chi(\alpha)|z\>\Big).
\]
Hence $|\alpha z\>=\chi(\alpha)|z\>$ and \gzit{equiv.sep} holds with
$\lambda_\alpha:=\chi(\alpha)$. Conversely, suppose that
\gzit{equiv.sep} holds for some complex number $\lambda_\alpha$.
Then, for $z,z'\in Z$, we get
\[
K(z',\alpha z)=\<z'|\alpha z\>=\<z'|\Big(\lambda_\alpha|z\>\Big)
=\lambda_\alpha\<z'|z\>=\lambda_\alpha K(z',z).
\]
This implies that $\alpha$ is a factorizing map with the factorization
constant $\chi(\alpha):=\lambda_\alpha$.
\epf

\begin{prop}~\\
(i) Every factorizing map $\alpha:Z\to Z$ satisfies
\lbeq{chi.bar}
K(\alpha z,z')=\ol{\chi(\alpha)}K(z,z') \for z,z'\in Z.
\eeq

(ii) Every factorizing map $\alpha$ with $\chi(\alpha)=1$
is coherent, with adjoint $1$.

(iii) Every factorizing map $\alpha:Z\to Z$ satisfies
\[
K(\alpha z,\alpha z')=|\chi(\alpha)|^2K(z,z')\Forall z,z'\in Z.
\]
(iv) Every factorizing map is homogeneous.
\end{prop}

\bepf
(i) and (ii) are straightforward.

(iii) Let $\alpha\in\Fac Z$ and $z,z'\in Z$. Then \gzit{chi.bar}
implies
\[
K(\alpha z,\alpha z')=\ol{\chi(\alpha)}K(z,\alpha z')
=\chi(\alpha)\ol{\chi(\alpha)}K(z,z')
=|\chi(\alpha)|^2K(z,z')
\]
(iv) Let $\alpha:Z\to Z$ be a factorizing map with the factorization
constant $\chi(\alpha)$. Suppose that $z,z'\in Z$ and
$\lambda\in\Cz^\*$ with $|z'\>=\lambda|z\>$. Then, for $z''\in Z$,
\[
\bary{lll}
\<z''|\alpha z'\>
&=&K(z'',\alpha z')=\chi(\alpha)K(z'',z')\\
&=&\chi(\alpha)\<z''|z'\>=\chi(\alpha)\lambda\<z''|z\>
=\lambda\<z''|\alpha z\>.
\eary
\]
Thus $|\alpha z'\>=\lambda|\alpha z\>$. Therefore, $m=1$ is a
multiplier for $\alpha$ and hence $\alpha$ is homogeneous.
\epf

We denote the set of all factorizing maps by $\Fac Z$ and the set of all
factorizing maps with nonzero factorization constants by $\Fac_\* Z$.
It is easy to check that any invertible factorizing map has a nonzero
factorization constant.

\begin{prop}
Let $Z$ be a coherent space. Then:

(i) The identity $1$ is a factorizing map with $\chi(1)=1$.

(ii) The composition of factorizing maps is factorizing.

(iii) Any adjoint $\alpha^*$ of a coherent and factorizing map $\alpha$
is factorizing with $\chi({\alpha^*})=\ol{\chi(\alpha)}$.

(iv) The inverse $\alpha^{-1}$ of any invertible factorizing map is
factorizing with $\chi(\alpha^{-1})=\chi(\alpha)^{-1}$.
\end{prop}

\bepf
(i) and (ii) are straightforward.

(iii) Let $\alpha:Z\to Z$ be a coherent and factorizing map and let
$\alpha^*$ be an adjoint for $\alpha$. Using \gzit{chi.bar} we find for
$z,z'\in Z$,
\[
K(\alpha^*z,z')=K(z,\alpha z')=\chi(\alpha)K(z,z').
\]
This implies that $\alpha^*$ is factorizing with
 $\chi({\alpha^*}):=\ol{\chi(\alpha)}$.

(iv) Let $\alpha\in\Fac Z$ be invertible with the inverse $\alpha^{-1}$.
Since $\chi(\alpha)\not=0$, for $z,z'\in Z$, we have
\[
K(\alpha^{-1}z,z')=\ol{\chi(\alpha^{-1})}K(\alpha\alpha^{-1}z,z')
=\ol{\chi(\alpha)^{-1}}K(z,z'),
\]
which implies that $\alpha^{-1}$ is factorizing with factorization 
constant $\chi(\alpha^{-1}):=\chi(\alpha)^{-1}$.
\epf

\begin{prop}
Let $Z$ be a coherent space. Then,

(i) $\Fac Z$ is a semigroup with identity.

(ii) $\Fac Z\cap\coh Z$ is $*$-semigroup.

(iii) The factorizing maps $\alpha$ with $\chi(\alpha)=1$ form a
subsemigroup
$\Fac_1(Z)$ of $\Fac Z$.

(iv) In the nondegenerate case, $\chi$ is an injective
multiplicative homomorphism into $\Cz$ and
$\Fac_1(Z)$ consists of the identity only.

(v) Each factorizing map $\alpha$ with $|\chi(\alpha)|=1$ preserves the
coherent product.
In particular, elements of $\Fac_1(Z)$ preserves the coherent product.
\end{prop}

\bepf
Straightforward.
\epf

\begin{thm}
Let $Z$ be a coherent space. Then $Z_{\times}:=(\Fac Z)\times Z$ with
the coherent product
\lbeq{K.times}
K_\times((\alpha,z);(\alpha',z')):=K(\alpha z,\alpha' z')
\Forall (\alpha,z),(\alpha',z')\in Z_\times
\eeq
is a coherent space.
\end{thm}
\bepf
Let $\alpha_1,..,\alpha_n\in\Fac Z$ and $z_1,..,z_n\in Z$.
Then, for all $c_1,...,c_n\in\Cz$, we have 
\[
\bary{lll}
\D\sum_{j,k}\ol{c_j}c_kK_\times((\alpha_j,z_j);(\alpha_k,z_k))
&=&\D\sum_{j,k}\ol{c_j}c_kK(\alpha_kz_j,\alpha_jz_k)\\
&=&\D\sum_{j,k}\ol{c_j}c_k\ol{\chi(\alpha_j)}\chi(\alpha_k)K(z_j,z_k)
\\
&=&\D\sum_{j,k}\ol{d_j}d_kK(z_j,z_k)\ge 0,
\eary
\]
where $d_\ell:=c_\ell\chi(\alpha_\ell)$ for $1\le \ell\le n$.
\epf

\begin{thm}\label{HA}
Let $Z$ be a coherent space. Then, for any $A:Z\to Z$ and
$f:Z\to\Cz^\*$, the map $\mathcal{H}_{(\alpha,A)}:PZ\to PZ$ defined 
on the projective extension $PZ$ via
\[
\mathcal{H}_{(f,A)}(\lambda,z):=(f(z)\lambda,Az)
\Forall (\lambda,z)\in PZ,
\]
is homogeneous in $\lambda$.
\end{thm}

\bepf
Let $\lambda,\lambda'\in\Cz^\*$ and $z\in Z$. Then, we have
\[
\mathcal{H}_{(f,A)}(\lambda'\lambda,z)=(f(z)\lambda'\lambda,Az)
=(\lambda'f(z)\lambda,Az)=\lambda'(f(z)\lambda,Az)
=\lambda'\mathcal{H}_{(f,A)}(\lambda,z).
\]
\epf

\begin{prop}\label{sep.proj.non.mult}
The factorizing maps on a projective and nondegenerate coherent
space of degree $e=\pm1$ are precisely the multiplication maps.
\end{prop}

\bepf
Clearly each multiplication map on a projective and non-degenerate
coherent space is factorizing. Conversely, let $Z$ be such a coherent
space and let $\alpha$ be a factorizing map with factorization constant
$\chi(\alpha)$. Then, for $z,z'\in Z$,
\[
K(z,\alpha z')=\chi(\alpha)K(z,z')=K(z,\chi(\alpha)^e z')
\]
since $\alpha$ is factorizing and $Z$ is projective.
Since $Z$ is nondegenerate we conclude $\alpha z=\chi(\alpha)^ez$.
\epf

For any coherent space $Z$, $PZ$ denotes the projective extension
defined in \cite[Proposition 5.8.5]{Neu.CQP}, with the same quantum
spaces as $Z$.

\begin{thm}\label{ST}
Let $Z$ be a coherent space, $S:Z\to Z$ be a factorizing map with
factorization constant $\chi(S)\in\Cz$. Then, the maps
$\mathcal{A}_S:PZ\to PZ$ and $\mathcal{B}_S:PZ\to PZ$ defined via
\[
\mathcal{A}_S(\lambda,z):=(\lambda,Sz)\Forall (\lambda,z)\in PZ,
\]
\[
\mathcal{B}_S(\lambda,z):=(\ol{\chi(S)}\lambda,z)\Forall (\lambda,z)
\in PZ,
\]
are coherent with $\mathcal{A}_S^*=\mathcal{B}_S$ and
$\mathcal{B}_S^*=\mathcal{A}_S$.
\end{thm}

\bepf
Let $(\lambda,z),(\lambda',z')\in PZ$. Then, we have
\[
\bary{lll}
K_{\rm pe}(\mathcal{A}_S(\lambda,z),(\lambda',z'))
&=&K_{\rm pe}((\lambda,Sz),(\lambda',z'))=\ol{\lambda}K(Sz,z')\lambda'\\
&=&\ol{\lambda}K(z,z')\ol{\chi(S)}\lambda'
=K_{\rm pe}((\lambda,z),\mathcal{B}_S(\lambda',z')).
\eary
\]
Thus, $\mathcal{A}_S$ is coherent with $\mathcal{A}_S^*=\mathcal{B}_S$.
This also implies that $\mathcal{B}_S$ is coherent with
$\mathcal{B}_S^*=\mathcal{A}_S$.
\epf

\begin{prop}
Let $Z$ be a coherent space. Then:

(i) The map $\mathcal{P}:\Cz\times\coh Z\to\coh PZ$ given by
$(\alpha,A)\to[\alpha,A]$ is an anti-homomorphism of
$*$-semigroups.

(ii) The map $\mathcal{A}:\Fac Z\to\coh PZ$ given by
$S\to\mathcal{A}_S$ is
a homomorphism of semigroups.

(iii) The map $\mathcal{B}:\Fac Z\to\coh PZ$ given by
$S\to\mathcal{B}_S$ is a homomorphism of semigroups.
\end{prop}

\bepf
(i) Let $(\alpha,A),(\beta,B)\in\Cz^\*\times\coh Z$. Then, for
$(\lambda,z)\in PZ$, we have
\[
\bary{lll}
(\alpha,A)\mathcal{P}_{(\beta,B)}(\lambda,z)
&=&\mathcal{P}_{(\beta,B)}(\alpha\lambda,Az)=(\beta\alpha\lambda,BAz)\\
&=&\mathcal{P}_{(\beta\alpha,BA)}(\lambda,z)
=\mathcal{P}_{(\beta,B)(\alpha,A)}(\lambda,z).
\eary
\]

(ii) Let $S,S'\in\Fac Z$. Then, for $(\lambda,z)\in PZ$, we have
\[
\mathcal{A}_{SS'}(\lambda,z)=(\lambda,SS'z)=\mathcal{A}_S(\lambda,S'z)
=\mathcal{A}_S\mathcal{A}_{S'}(\lambda,z).
\]
(iii) Let $S,S'\in\Fac Z$. Then, for $(\lambda,z)\in PZ$, we have
\[
\mathcal{B}_{SS'}(\lambda,z)=(\chi(SS')\lambda,z)
=(\chi(S)\chi(S')\lambda,z)
=\mathcal{B}_S(\chi(S')\lambda,z)
=\mathcal{B}_S\mathcal{B}_{S'}(\lambda,z).
\]
\epf

\begin{cor}\label{sep.npz}
Let $Z$ be a coherent space. Then
\[
\Fac\ [PZ]\cong\Fac P[Z]\cong\Cz^\*.
\]
In particular, the map $\chi:\Fac\ [PZ]\cong\Fac P[Z]\to\Cz^\*$ is a
group isomorphism.
\end{cor}

A map $A:Z\to Z$ is called \bfi{strongly homogeneous} if
$A\alpha=\alpha A$ for all factorizing maps $\alpha\in\Fac Z$. We write
$\Hom_s Z$ for the set of all strongly homogeneous maps over $Z$.
It can be readily checked that $\Fac Z\subseteq \Hom_s Z$ and
$\Hom_s Z\subseteq\Hom(Z)$.

A function $f:Z\to\Cz$, or a kernel $X:Z\times Z\to \Cz$ is called
\bfi{strongly homogeneous} if
\[
f(\alpha z)=f(z) \for \alpha\in \Fac Z,~z\in Z,
\]
or
\[
X(\alpha z,\alpha'z')=X(z,z')
 \for \alpha,\alpha'\in \Fac Z,~z,z'\in Z,
\]
respectively.

\begin{prop}
Let $Z$ be a coherent space. 

(i) Any adjoint of a coherent and strongly homogeneous map is
homogeneous.

(ii) The set $\coh Z\cap{\rm Hom}(Z)$ is $*$-subsemigroup of
$\coh Z$.
\end{prop}

\bepf
(i) Let $A:Z\to Z$ be a strongly homogeneous coherent map with an
adjoint $A^*$. Then, for all $\alpha\in\Fac Z$, we have
\[
K(A^*(\alpha z),z')=K(\alpha z,Az')=\chi(\alpha)K(z,Az')
=\chi(\alpha)K(A^*z,z')
=K(\alpha A^*z,z'),
\]
for all $z,z'\in Z$. Thus, $A^*$ is strongly homogeneous.

(ii) is straightforward.
\epf

\begin{prop}\label{main.hom.p}
Let $Z$ be a nondegenerate coherent space. Then,

(i) each coherent map is strongly homogeneous.

(ii) $\Fac Z$ is in the center of $\coh Z$.

(iii) For $z\in Z$, $\alpha\in\Fac Z$, and $A\in\coh Z$ we have
$|A\alpha z\>=\chi(\alpha)|Az\>$.
\end{prop}
\bepf
(i) Let $A:Z\to Z$ be a coherent map. Then, for all $z,z'\in Z$ and
$\alpha\in\Fac Z$,
we have
\[
K(A\alpha z,z')=K(\alpha z,A^*z')=\ol{\chi(\alpha)}K(z,A^*z')
=\ol{\chi(\alpha)}K(Az,z')=K(\alpha Az,z').
\]
Since $K$ is nondegenerate over $Z$, we get
$A\circ\alpha=\alpha\circ A$ for all $\alpha\in\Fac Z$.

(ii) Let $\alpha\in\Fac Z$ with the factorization constant 
$\chi(\alpha)$.
Also, let $A\in\coh Z$ be given. Using (i), $A$ is strongly homogeneous
as well. Thus, by definition of strongly homogeneous we have
$A\alpha=\alpha A$. Hence $\alpha$ belongs to the center of $\coh Z$.

(iii) By (ii), $|A\alpha z\>=|\alpha Az\>=\chi(\alpha)|Az\>$.
\epf

\begin{prop}
Let $Z$ be a coherent space and $z,z'\in Z$.
If there exists a factorizing map $\alpha\in\Fac Z$ such that
$\alpha z=z'$ then the coherent states $|z\>,|z'\>$
are parallel. In this case, we have $|z'\>=\chi(\alpha)|z\>$.
\end{prop}

\bepf
Suppose that there exists a factorizing map $\alpha\in\Fac Z$ such that
$\alpha z=z'$. Then, for $w\in Z$, we have
\[
\<w|z'\>=K(w,z')=K(w,\alpha z)=\chi(\alpha)K(w,z)
=\chi(\alpha)\<w|z\>.
\]
Thus we get $|z\>=\chi(\alpha)|z'\>$.
\epf

\begin{rem}
If $Z$ is a projective and nondegenerate coherent space then
$\Hom Z=\Hom_s Z$. Indeed, a function $f:Z\to\Cz$, or a kernel
$X:Z\times Z\to \Cz$ is homogeneous
iff
\[
f(\alpha z)=f(z) \for \alpha\in \Cz^\*,~z\in Z,
\]
or
\[
X(\alpha z,\alpha'z')=X(z,z')
 \for \alpha,\alpha'\in \Cz^\*,~z,z'\in Z,
\]
respectively.
\end{rem}

The next result shows that each coherent map over a projective coherent
space is automatically homogeneous as well.

\begin{cor}\label{cor.non.cent}
Let $Z$ be a projective and nondegenerate coherent space.
Then,

(i) every coherent map is homogeneous.

(ii) $\Cz^\*$ is in the center
of $\coh Z$.
\end{cor}

\bepf
The results follow directly from Propositions \ref{main.hom.p} and
 \ref{sep.proj.non.mult}.
\epf

\begin{cor}
Let $Z$ be a coherent space. Then

(i) $\coh\ [PZ] \subseteq\Hom\ [PZ]$ and $\coh P[Z]\subseteq\Hom P[Z]$.

(ii) $\Fac\ [PZ]$ is in the center of $\coh\ [PZ]$.

(iii) $\Fac P[Z]$ is in the center of $\coh P[Z]$.
\end{cor}

\bepf
Apply Corollary \ref{cor.non.cent} to the projective and
non-degenerate spaces $Z':=[PZ]$ and $Z'':=P[Z]$.
\epf

\begin{prop}\label{main.sep}
Let $Z$ be a coherent space and $z\in Z$. Then

(i) For $\alpha\in\Fac Z$ and $A\in\Hom_s Z$ we have
\[
|A\alpha z\>=\chi(\alpha)|Az\>.
\]
(ii) For $A\in\coh Z$ and $\alpha\in\Fac Z$ we have
\[
|A\alpha z\>=\chi(\alpha)\Gamma(A)|z\>=|\alpha Az\>.
\]
\end{prop}
\bepf
Straightforward.
\epf

\section{Slender coherent spaces}\label{s.slender}

In this section we prove quantization theorems for the class of slender
coherent spaces for which the admissibility of functions, defined 
in Section \ref{s.adFunc}, is particularly easy to check. For slender 
coherent spaces, many operators on a quantum space have a simple 
description in terms of normal kernels. These generalize the normal 
ordering of operators familiar from quantum field theory.

\subsection{Slender coherent spaces}

\begin{prop}\label{p.all}
For a coherent space $Z$, the following are equivalent:

(i) Every function $f:Z\to\Cz$ is admissible.

(ii) Every finite set of distinct coherent states is linearly 
independent.
\end{prop}

\bepf
If any finite set of distinct coherent states is linearly independent
then the hypothesis of \gzit{e.fIffSym} implies that all $c_k$ vanish. 
Thus each function $f:Z\to\Cz$ is admissible. Hence $\Az(Z)=\Cz^Z$.
Conversely, suppose that $\Az(Z)=\Cz^Z$, and
$\D\sum c_\ell|z_\ell\>=0$ with distinct $z_\ell$. Then every
$f=\delta_{z_k}$ is admissible and 
\[
0=\sum\ol c_\ell f(\ol z_\ell)
=\sum\ol c_\ell \delta_{z_k}(z_\ell)=\ol{c_k},
\]
which implies that $c_k=0$. This holds for all $k$, whence any finite
set of distinct coherent states is linearly independent.
\epf

The most interesting cases are covered by a slightly more general
class of coherent spaces.
We call a coherent space \bfi{slender} if any finite set of linearly
dependent, nonzero coherent states in a quantum space $\Qz(Z)$ of $Z$
contains two parallel coherent states. Clearly, every subset of a
slender coherent space is again a slender coherent space.

\begin{prop}
Let $S$ be a subset of the Euclidean space $\Hz$ such that any two
elements of $S$ are linearly independent. Then the set
$Z=\Cz^\*\times S$ with the coherent product
\[
K((\lambda,s);(\lambda',s')):=\ol{\lambda}\lambda's^*s'
\Forall (\lambda,s),(\lambda',s')\in Z
\]
and scalar multiplication $\alpha(\lambda,s):=(\alpha\lambda,s)$
is a slender, projective coherent space of degree 1.
\end{prop}

\bepf
It is easy to see that $\Qz(Z):=\Span S$ is a quantum space of
$Z$. Let the $z_k\in Z$ be such that $\D\sum c_k|z_k\>=0$ with
$c_k\not=0$ for all $k$. We then have
$z_k=(\lambda_k,z_k')$ with $\lambda_k\in\Cz^\*$ and $z_k'\in S$, hence
\[
\sum c_k\lambda_k z_k'=\sum c_k|(\lambda_k,z_k')\>
=\sum c_k|z_k\>=0.
\]
But the $z_k'$ are linearly independent, hence $c_k\lambda_k=0$ for all
$k$, and since $\lambda_k\ne 0$, all $c_k$ vanish. Thus $Z$ is slender.
Projectivity is obvious.
\epf

Thus slender coherent spaces are very abundant. However, proving
slenderness for a {\em given} coherent space is a nontrivial matter once
$Z$ contains infinitely many elements.

\begin{thm} \label{t.slenderMob}
The M\"obius space defined in Example \ref{ex.Moebius}
is a slender coherent space.
\end{thm}

\bepf
Suppose that there is a nontrivial finite linear dependence 
$\D\sum c_k|z_k\>=0$ such that no two $|z_k\>$ are parallel. 
By definition of $Z$, the numbers $\mu_k:=z_{k2}/z_{k1}$ satisfy
$|\mu_k|<1$. Moreover, $z_k=z_{k1}{1\choose \mu_k}$, and since $Z$ is 
projective of degree $-1$,
\[
|z_k\>=z_{k1}^{-1}\Big|{1\choose \mu_k}\Big\>.
\]
Since no two $|z_k\>$ are parallel, the $\mu_k$ are distinct. Since 
$x={1\choose \mu}\in Z$ for $|\mu|<1$, we have
\[
0=\<\ol x|\sum c_k|z_k\>=\sum c_kK(\ol x,z_k)
=\sum \frac{c_k}{x_1z_{k1}-x_2z_{k2}}
=\sum \frac{c_kz_{k1}^{-1}}{1-\mu\mu_k} \for |\mu|<1.
\]
The right hand side is the partial fraction decomposition of a
rational function of $\mu$ vanishing in an open set. Since the
partial fraction decomposition is unique, each term vanishes.
Therefore $c_kz_{k1}^{-1}=0$ for all $k$, which implies that all
$c_k$ vanish, contradiction. Thus $Z$ is slender.
\epf

\begin{prop}~\\
(i) A projective coherent space is slender iff
$\D\sum_{k\in I} |z_k\>=0$ implies that there exist distinct $j,k\in I$
such that $|z_k\>=\alpha|z_j\>$ for some $\alpha\in\Cz$.

(ii) A nondegenerate projective coherent space is slender iff
$\D\sum_{k\in I} |z_k\>=0$ implies that there exist distinct $j,k\in I$
such that $z_k=\alpha z_j$ for some $\alpha\in\Cz$.

(iii) A coherent space $Z$ is slender iff its projective extension $PZ$
is slender.
\end{prop}

\bepf
In the projective case, $\D\sum \alpha_k|z_k\>=0$ implies
$\D\sum |\beta_kz_k\>=0$ with $\beta_k:=\alpha_k^{1/e}$. Thus we may
assume w.l.o.g. that the linear combination in the definition of
slender is a sum. Hence (i) holds. (ii) is straightforward.

(iii) Let $Z$ be a slender coherent space with a quantum space $\Qz(Z)$,
and let $PZ$ be a projective extension of $Z$ of degree $e$ with the
same quantum space $\Qz(PZ)=\Qz(Z)$.
Let $\D\sum_{k}|(\lambda_k,z_k)\>=0$ in $\Qz(PZ)$.
Then $\D\sum_k\lambda_k^e|z_k\>=0$ in $\Qz(Z)$, and we may
assume that the sum extends only over the nonzero $\lambda_k$.
Since $Z$ is slender, there exists distinct $j,k$ with $\lambda_k\ne 0$
such that $|z_j\>=\alpha|z_k\>$ for some $\alpha\in\Cz$. But then
\[
|(\lambda_j,z_j)\>=\lambda_j^e|z_j\>=\lambda_j^e\alpha|z_k\>
=\Big(\frac{\lambda_j}{\lambda_k}\Big)^e\alpha|(\lambda_k,z_k)\>.
\]
Thus $PZ$ is slender. The converse is obvious.
\epf

\begin{prop}\label{p.slender}
Let $Z$ be a slender coherent space and let $\Qz(Z)$ be a quantum space
of $Z$. Let $I$ be a finite index set. If the $z_k\in Z$ ($k\in I$)
satisfy $\D\sum_{k\in I}c_k|z_k\>=0$ then there is a partition of $I$
into nonempty subsets $I_t$ ($t\in T$) such that $k\in I_t$ implies
$|z_k\>=\alpha_k|z_t\>$ with $\D\sum_{k\in I_t}c_k\alpha_k=0$
for all $t\in T$.
\end{prop}

\bepf
It is easy to see that it is enough to consider the case where none of
the $c_k|z_k\>$ vanishes, as the general case can be reduced to this
case by removing zero contributions to the sum.
Let $T$ be a maximal subset of $I$ with the property that no two
coherent states $|z_t\>$ are multiples of each other. For each $t\in T$,
let $I_t$ be the set of $k\in I$ such that $|z_k\>$ is a multiple of
$|z_t\>$, say, $|z_k\>=\alpha_k|z_t\>$. Then the $I_t$ ($t\in T$) form
a partition of $I$. If we define for $t\in T$ the numbers
\[
a_t:=\sum_{k\in I_t}c_k\alpha_k
\]
we have
\[
\sum_{t\in T}a_t|z_t\>
=\sum_{t\in T}\Big(\sum_{k\in I_t}c_k\alpha_k\Big)|z_t\>
=\sum_{k\in I}c_k|z_k\>=0.
\]
Since $Z$ is a slender coherent space and no two of the $|z_t\>$
($t\in T$) are parallel, the $|z_t\>$ ($t\in T$) are linearly
independent. We conclude that all $a_t$ vanish. Therefore
$\D\sum_{k\in I_t}c_k\alpha_k=0$ for all $t\in T$.
\epf

\begin{cor}\label{main.slender.proj}
Let $Z$ be a slender coherent space, projective of degree $e$ and let
$\Qz(Z)$ be a quantum space of $Z$. Let $I$ be a finite index set.
If $z_k\in Z$ ($k\in I$)
satisfies $\D\sum_{k\in I}|z_k\>=0$ then there is a partition of $I$
into nonempty subsets $I_t$ ($t\in T$) such that $k\in I_t$ implies
$z_k=\alpha_k z_t$ with $\D\sum_{k\in I_t}\alpha_k^e=0$,
for all $t\in T$.
\end{cor}

\subsection{Quantization theorems}label{ss.normal}

\begin{thm}\label{main.quant.slender}
Let $Z$ be a slender coherent space and suppose that $m:Z\to \Cz$
is a multiplier map for the map $A:Z\to Z$. Then there exists a unique
linear operator $\Gamma_m(A):\Qz(Z)\to \Qz(Z)$, the \bfi{quantization}
of $A$ relative to $m$, such that
\[
\Gamma_m(A)|z\>=m(z)|Az\>\Forall z\in Z.
\]
\end{thm}

\bepf
Let $m:Z\to\Cz$ be a multiplier for the map $A:Z\to Z$.
We then define $\Gamma_m(A):\Qz(Z)\to\Qz(Z)$ by
\[
\Gamma_m(A)\Big(\sum_kc_k|z_k\>\Big):=\sum_kc_km(z_k)|Az_k\>
\Forall \sum_k c_k|z_k\>\in\Qz(Z).
\]
Let $\D\sum c_k|z_k\>=0$. Then, we have $\D\sum|c_kz_k\>=0$. Hence,
using Proposition \ref{p.slender}, there is a partition of
$I:=\{k:c_k\not=0\}$
into nonempty subsets $I_t$ ($t\in T$) such that $k\in I_t$ implies
$|z_k\>=\alpha_k|z_t\>$ with $\D\sum_{k\in I_t}c_k\alpha_k=0$,
for all $t\in T$. Since $\alpha$ is a multiplier for $A$, we have for
$t\in T$ and $k\in I_t$,

\lbeq{mult.loc}
m(z_k)|Az_k\>=\alpha_km(z_t)|Az_t\>.
\eeq
Thus, using (\ref{mult.loc}), we get
\[
\bary{lll}
\D\sum_kc_km(z_k)|Az_k\>
&=&\D\sum_{t\in T}\sum_{k\in I_t}c_km(z_k)|Az_k\>
=\sum_{t\in T}\sum_{k\in I_t}c_k\alpha_km(z_t)|Az_t\>\\
&=&\D\sum_{t\in T}\Big(\sum_{k\in I_t}c_k\alpha_k\Big)m(z_t)|Az_t\>=0.
\eary
\]
Therefore, $\Gamma_m(A):\Qz(Z)\to\Qz(Z)$ is a well-defined linear map.
In particular, we have
\[
 \Gamma_m(A)|z\>=m(z)|Az\> \for z\in Z.
\]
\epf

\begin{cor}\label{t.quant1}
Let $Z$ be a slender, projective, and non-degenerate coherent space,
and let $\Qz(Z)$ be a  quantum space of $Z$.
Then for every homogeneous map $A:Z\to Z$, there is a unique linear
operator $\Gamma(A):\Qz(Z)\to\Qz(Z)$, the \bfi{quantization} of $A$,
such that
\lbeq{Gamma.1.A}
\Gamma(A)|z\>=|Az\>   \for z\in Z.
\eeq
\end{cor}
\bepf
We define $\Gamma(A):\Qz(Z)\to\Qz(Z)$ by $\Gamma(A):=\Gamma_1(A)$.
Then, $\Gamma(A)$ satisfies (\ref{Gamma.1.A}).
\epf

Note that the results just proved assume that $Z$ is slender but need
minimal assumption about $A$ In contrast, Theorem \ref{t.Gamma} holds 
for arbitrary coherent spaces, but it assumes that $A$ is a coherent 
map. The simple relationship \gzit{e.GammaStar} between $\Gamma(A)^*$ 
and $\Gamma(A^*)$, valid for coherent maps $A$, does not generalize to 
the situation discussed in the present section.

\begin{prop}
Let $Z$ be a slender, projective and non-degenerate coherent space.
The quantization map $\Gamma:\Hom Z\to\Lin\Qz(Z)$ is
a semigroup homomorphism,
\lbeq{e.GammaAB}
\Gamma(AB)=\Gamma(A)\Gamma(B)\for A,B\in\Hom Z.
\eeq
\end{prop}

\bepf
It is straightforward to check that $AB\in\Hom Z$.
By Theorem \ref{t.quant1},
\[
\Gamma(AB)|z\>=|ABz\>=\Gamma(A)|Bz\>=\Gamma(A)\Gamma(B)|z\>
\Forall z\in Z.
\]
Thus \gzit{e.GammaAB} holds.
\epf

\begin{thm}\label{t.quant3}
Let $Z$ be a slender coherent space and let $\Qz(Z)$ be a quantum space
of $Z$. Then for every homogeneous function $m:Z\to\Cz$ there is a
unique linear operator $\a(m):\Qz(Z)\to\Qz(Z)$ such that
\lbeq{e.am}
\a(m)|z\>=m(z)|z\> \for z\in Z.
\eeq
\end{thm}

\bepf
We define $\a(m):\Qz(Z)\to\Qz(Z)$ by $\a(m):=\Gamma_m(1)$. Then
\gzit{e.am} follows easily.
\epf

This generalizes the property of traditional coherent states to be
eigenstates of annihilator operators. Indeed, in the special case
of Klauder spaces treated in Subsection \ref{ss.lowering}, the $\a(m)$
are found to be the smeared annihilator operators acting on a Fock
space.

$\a(m)$ is a linear function of $m$. To preserve this property in the
adjoint, we define
\lbeq{e.asm}
\a^*(m):=\a(\ol m)^*,
\eeq
the analogues of smeared creation operators. Here $\ol m$ is the
function defined by 
\[
\ol m(z):=\ol{m(z)},
\]
which is homogeneous since $|z'\>=\lambda |z\>$ implies $m(z')= m(z)$ 
by homogeneity of $m$, hence $\ol m(z')=\ol{m(z')}=\ol{m(z)}=\ol m(z)$.

\subsection{Normal ordering}\label{ss.normal}

A kernel $X:Z\times Z\to\Cz$ on a coherent space $Z$ is called 
\bfi{homogeneous} if, for all $z\in Z$, the functions $X(\cdot,z)$ and 
$X(z,\cdot)$ are homogeneous in the sense defined in Subsection 
\ref{ss.hom}. The homogeneous kernels on $Z$ form a vector space 
$\Xz(Z)$.

\begin{thm}\label{t.normalOrder}
Let $Z$ be a slender coherent space and $\Qz(Z)$ be a quantum space of
$Z$. Then, for every homogeneous kernel $X$, there is a unique linear 
operator $N(X)$ from $\Qz(Z)$ to $\Qz^\*(Z)$, called the 
\bfi{normal ordering} of $X$, such that
\lbeq{e.normal}
\<z|N(X)|z'\>=X(z,z')K(z,z') \for z,z'\in Z.
\eeq
Equivalently, $N(X)$ defines a Hermitian form on $\Qz(Z)$.
\end{thm}

The name indicates a relation to the normal ordering prescription for 
operators in quantum field theory; cf. Theorem \ref{t.normalOrd} below. 

\bepf
This follows from Theorem \ref{t.opExist}(ii) and slenderness. To see
this, we define, for any two vectors $\phi=\D\sum_kc_k'|z_k'\>$ and
$\psi=\D\sum_\ell c_\ell|z_\ell\>$ from $\Qz(Z)$, the complex number
\[
(\psi,\phi)_X
:=\sum_\ell\sum_k\ol{c_\ell}{c_k'}X(z_\ell,z_k')K(z_\ell,z'_k).
\]
We first claim that $(\psi,\phi)\to(\psi,\phi)_X$ is well-defined.
Because $(.,.)_X$ is a Hermitian form in the $c_\ell$ and the $c_k'$,
it is enough to show that $\phi=0$ implies $(\psi,\phi)_X=0$.
By Proposition \ref{p.slender}, if $\phi=\D\sum_kc_k'|z_k'\>=0$, there
is a partition of $I:=\{k:c_k'\not=0\}$ into nonempty subsets $I_t$
($t\in T$) such that $k\in I_t$ implies $|z_k'\>=\alpha_k|z_t'\>$ with
$\D\sum_{k\in I_t}c_k'\alpha_k=0$ for all $t\in T$. Using the
homogeneity assumption of $X(z_\ell,\cdot)$ we find for each $t\in T$
and each $k\in I_t$,
\[
X(z_\ell,z_k')K(z_\ell,z_k')=\alpha_kX(z_\ell,z_t')K(z_\ell,z_t').
\]
Therefore
\[
\bary{lll}
\D\sum_\ell\sum_k\ol{c_\ell}{c_k'}X(z_\ell,z_k')K(z_\ell,z'_k)
&=&\D\sum_\ell\sum_{t\in T}
     \sum_{k\in I_t}\ol{c_\ell}{c_k'}X(z_\ell,z_k')K(z_\ell,z'_k)\\[5mm]
&=&\D\sum_\ell\sum_{t\in T}
\sum_{k\in I_t}\ol{c_\ell}{c_k'}\alpha_kX(z_\ell,z_t')K(z_\ell,z'_t)
\\[5mm]
&=&\D\sum_\ell\sum_{t\in T}\ol{c_\ell}
\Big(\sum_{k\in I_t}{c_k'}\alpha_k\Big)X(z_\ell,z_t')K(z_\ell,z'_t)=0.
\eary
\]
Using the homogeneity assumption of $X(\cdot,z_k')$, a similar argument
shows that if $\psi=\sum c_\ell|z_\ell\>=0$ then $(\psi,\phi)_X=0$.
Hence, $(\psi,\phi)\to (\psi,\phi)_X$ defines a well-defined
Hermitian form on $\Qz(Z)$.
\epf

\begin{prop}\label{p.nowhere}
Let $Z$ be a slender coherent space whose coherent product vanishes
nowhere. Then any linear operator $\X:\Qz(Z)\to\Qz(Z)^\*$
is the normal ordering of a unique homogeneous kernel $X$.
\end{prop}

\bepf
The kernel $X$ defined by
\[
X(z,z'):=\frac{\<z|\X|z'\>}{K(z,z')}
\]
is homogeneous and satisfies \gzit{e.normal}. Hence $N(X)=\X$ by
Theorem \ref{t.normalOrder}.
\epf

\begin{prop}
If $A:Z\to Z$ is coherent, homogeneous and invertible then
$\Gamma(A)$ is invertible, and for every homogeneous kernel $X$,
the kernel $AX$  defined by
\lbeq{e.AX}
AX(z,z'):=X(A^*z,A^{-1}z') \for z,z'\in Z
\eeq
is homogeneous, and
\[
N(AX)=\Gamma(A)N(X)\Gamma(A)^{-1}.
\]
Moreover, if  $B:Z\to Z$ is also coherent,  homogeneous and invertible
then
\[
(AB)X=A(BX).
\]
\end{prop}

\bepf
This follows from \gzit{e.AX} since
\[
\bary{lll}
\<z|N(AX)\Gamma(A)|z'\>&=&\<z|N(AX)|Az'\>=AX(z,Az')K(z,Az')\\
&=&X(A^*z,z')K(A^*z,z')=\<A^*z|N(X)|z'\>=\<z|\Gamma(A)N(X)|z'\>
\eary
\]
and
\[
\bary{lll}
(AB)X(z,z')&=&X((AB)^*z,(AB)^{-1}z')=X(B^*A^*z,B^{-1}A^{-1}z')\\
&=&BX(A^*z,A^{-1}z')=A(BX)(z,z').
\eary
\]
\epf

Define for $f,g:Z\to\Cz$,
\[
(fX)(z,z'):=f(z)X(z,z'),~~~(Xf)(z,z'):=X(z,z')f(z'),
\]
Then
\[
(fX)^*=X^*f^*,~~~(Xf)^*=f^*X^*.
\]
We write $\lambda$ for a \bfi{constant kernel} with constant value
$\lambda\in \Cz$. Note that $f1$ and $1f$ are different homogeneous 
kernels!

\begin{prop}~\\
(i) The normal ordering operator $N:\Xz(Z)\to\Linx\Qz(Z)$ is linear.

(ii) If $X$ is homogeneous then $X^*$ is homogeneous and
\[
N(X^*)=N(X)^*.
\]
(iii) Any constant kernel $\lambda$ is homogeneous, and 
$N(\lambda)=\lambda$.

(iv) If $N$ is homogeneous then $mNm'$ is homogeneous for all 
homogeneous $m,m'$, and
\lbeq{e.NmXm'}
N(mXm')=\a^*(m)N(X)\a(m').
\eeq
(v) If $X_\ell\to X$ pointwise and all $X_\ell$ are homogeneous then 
$X$ is homogeneous, and
\[
 N(X_\ell)\to N(X).
\]
\end{prop}

\bepf
Statements (i)--(iii) are straightforward.

(iv) Using \gzit{e.am} and \gzit{e.asm}, we find
\[
\bary{lll}
\<z|\a^*(m)N(X)\a(m')|z'\>
&=& \<z|\a(m^*)^*N(X)\a(m')|z'\>=\<z|m(z)N(X)m'(z')|z'\>\\
&=& m(z)\<z|N(X)|z'\>m'(z')= m(z)X(z,z')K(z,z')m'(z')\\
&=&(mXm')(z,z')K(z,z').
\eary
\]
(v) Let $X_\ell\to X$ pointwise with all $X_\ell$ homogeneous. Then $X$ 
is homogeneous. Indeed, for $z,z'\in Z$ and $c,c'\in\Cz$, we have
\[
X(cz,c'z')=\lim_\ell X_\ell(cz,c'z')=\lim_\ell X_\ell(z,z')=X(z,z').
\]
We then have
\[
\lim_\ell\<z|N(X_\ell)|z'\>=\Big(\lim_\ell X_\ell(z,z')\Big)K(z,z')
=X(z,z')K(z,z')=\<z|N(X)|z'\>,
\]
for all $z,z\in Z$.
\epf

\begin{thm}
Let $Z$ be a slender coherent space. Let $S$ be a set, $d\mu$ a measure
on $S$. Suppose that the $f_\ell,g_\ell:S\times Z\to \Cz$ are
measurable in the first argument and homogeneous in the second argument,
and
\[
X(z,z'):=\lim_\ell \int d\mu(s)g_\ell(s,z)f_\ell(s,z')
\]
exists for all $z,z'\in Z$. Then $X$ is homogeneous and, with notation 
as in \gzit{e.am},
\lbeq{e.NX}
N(X)=\lim_\ell \int d\mu(s)\a(g_\ell(s,\cdot))^*\a(f_\ell(s,\cdot)).
\eeq
\end{thm}

\bepf
Let $z,z'\in Z$ and $\alpha\in\Cz$ such that $|z'\>=\alpha|z\>$. Using 
the homogeneity of the $f_\ell$, we find for each $w\in Z$
\[
\bary{lll}
X(w,z')|z'\>
&=&\D\lim_\ell \int d\mu(s)g_\ell(s,w)f_\ell(s,z')|z'\>\\[4mm]
&=&\alpha\D\lim_\ell \int d\mu(s)g_\ell(s,w)f_\ell(s,z)|z\>
=\alpha X(w,z)|z\>.
\eary
\]
This implies that $X(w,\cdot)$ is a homogeneous function. A similar
argument, using homogeneity assumption of each $g_\ell$, guarantees
that $X(\cdot,w)$ is a homogeneous function as well. Thus $X$ is 
homogeneous and $N(X)$ is defined. Now for $z,z'\in Z$, 
\[
\bary{lll}
\<z|\D\lim_\ell \int d\mu(s)
           \a(g_\ell(s,\cdot))^*\a(f_\ell(s,\cdot))|z'\>
&=&\D\lim_\ell\int d\mu(s)
           \Big(\a(g_\ell(s,\cdot))^*\<z|\Big)\a(f_\ell(s,\cdot))|z\>\\
&=&\D\lim_\ell \int d\mu(s)g_\ell(s,z)f_\ell(s,z')K(z,z')\\
&=&\D X(z,z')K(z,z')=\<z|N(X)|z'\>.
\eary
\]
Thus \gzit{e.NX} holds by Theorem \ref{t.normalOrder}.
\epf

\newpage
\section{Klauder spaces and bosonic Fock spaces}\label{s.Klauder}

\subsection{Klauder spaces}

We recall from \sca{Neumaier} \cite[Example 5.2.2]{Neu.CQP} that the
\bfi{Klauder space} $Kl(V)$ over the Euclidean space $V$ is defined by
the set $Z=\Cz\times V$ of pairs
\[
z:=[z_0,\z]\in \Cz\times V
\]
with the coherent product
\lbeq{e.Kosc}
K(z,z'):=e^{\ol z_0 +z_0'+\z^*\z'}.
\eeq
Klauder spaces are degenerate since
\[
|[z_0+2\pi i k,\z]\>=|[z_0,\z]\> \for k\in\Zz.
\]
Thus it is enough to specify $z_0$ modulo $2\pi i$.

\begin{prop}
With the scalar multiplication
\[
\alpha[z_0,\z]:=[z_0+\log \alpha,\z],
\]
using an arbitrary but fixed branch of $\log$, Klauder spaces are
projective of degree 1. The factorizing maps are precisely the
multiplication maps $z\to \alpha z$, with $\chi(\alpha)=\alpha$.
\end{prop}

\bepf
Using the definition of the scalar multiplication, one finds
$K(z,\lambda z') = \lambda K(z,z')$. The second statement can be
verified directly; Proposition \ref{sep.proj.non.mult} is not
applicable.
\epf

\begin{thm}
Klauder spaces are slender.
\end{thm}

\bepf
Suppose that there is a nontrivial finite linear dependence 
$\D\sum c_k|z_k\>=0$ such that no two $|z_k\>$ are parallel. 
Since $z_0$ only contributes a scalar facto to $|z\>$, we may assume
w.l.o.g. that $z_k=[0,\z_k]$ and conclude that the $\z_k$ are distinct.
Now let $v\in V$ and $z=[0,nv]$ for some nonnegative integer $n$.
Then, with $\xi_k:=e^{v^*\z_k}$,
\[
0=\<z|\sum c_k|z_k\>=\sum c_k \<z|z_k\>=\sum c_k e^{n v^*\z_k}
=\sum c_k \xi_k^n \for n=0,1,2,\ldots.
\]
Since the sum has finitely many terms only, we find a homogeneous linear
system with a Vandermonde coefficient matrix having a nontrivial
solution. So the matrix is singular, and we conclude that two of the
$\xi_k$ must be identical. Thus for every $v\in V$ there are
indices $j<k$ such that $e^{v^*\z_j}=e^{v^*\z_k}$, hence, with
$z_{jk}:=z_j-z_k\ne 0$,
\[
v^*z_{jk}\equiv 0\mod 2\pi i.
\]
Now let $u\in V$. If $u^*z_{jk}\ne 0$ for all $j<k$
then picking $v=\lambda u$ with sufficiently many different
$\lambda\in\Rz$ gives a contradiction.
Thus for every $u\in V$  there are indices $j<k$ such that
\lbeq{e.eq}
v^*z_{jk}=0.
\eeq
Since $u\in V$ was arbitrary and the $z_{jk}$ are nonzero, this
implies that $V$ is the union of finitely many hyperplanes \gzit{e.eq},
which is impossible. 

Therefore the assumed nontrivial finite linear dependence does not 
exist. This proves that the Klauder space $Z=Kl(V)$ is slender. 
\epf

\subsection{Oscillator groups}

Klauder spaces have a large semigroup of coherent maps, which
contains a large unitary subgroup. We call $\A\in\Linx V$ 
\bfi{horizontal} if both $A$ and $A^*$ map $V$ into $V$ and $\ol V$ 
into $\ol V$. We write $\Hor V$ for the vector space of all horizontal 
linear maps $\A\in\Linx V$. The \bfi{oscillator semigroup} over $V$ is 
the semigroup $Os(V)$ of matrices
\[
A=[\rho,p,q,\A]:=\pmatrix{1 & p^* & \rho \cr 0 & \A & q \cr 0 & 0 & 1}
\in \Lin(\Cz\times V\times \Cz)
\]
with  $\rho\in \Cz$, $p,q\in V$, and $\A\in\Hor V$. One
easily verifies the formula for the product
\lbeq{e.oscProd}
[\rho,p,q,\A][\rho',p',q',\A']
    =[\rho'+\rho+p^*q',\A'^*p+p',q+\A q',\A\A']
\eeq
and the identity $1=[0,0,0,1]$. Writing
\[
[\A]:=[0,0,0,\A]
\]
we find
\lbeq{e.BOs}
[\B][\alpha,p,q,\A][\B']=[\alpha,\B'^*p,\B q,\B\A\B'].
\eeq
$Os(V)$ turns elements $z\in Z$ written in the projective form
\[
z=[z_0,\z]=\pmatrix{z_0 \cr \z \cr 1}\in \Cz^\*\times V\times \Cz
\]
into elements the same form, corresponding to the action of $Os(V)$ on
$[z_0,\z]\in Kl(V)$ as
\lbeq{e.oscCoh}
[\rho,p,q,\A][z_0,\z]:=[\rho+z_0+p^*\z,q+\A\z].
\eeq

\begin{prop}
$Os(V)$ is a *-semigroup of coherent maps of $Kl(V)$, with adjoints
defined by
\lbeq{e.oscAdj}
[\rho,p,q,\A]^*=[\ol\rho,q,p,\A^*].
\eeq
\end{prop}

\bepf
We have
\[
\bary{lll}
K(Az,z')&=&K([\rho,p,q,\A]z,z')
 =e^{\ol{\rho+z_0+p^*\z} +z_0'+(q+\A\z)^*\z'}\\[3mm]
&=&e^{\ol z_0+\ol\rho+z_0'+q^*\z' +\z^*(p+\A^*\z')}
=K(z,[\ol\rho,q,p,\A^*]z').
\eary
\]
Hence the elements of $Os(V)$ are coherent maps, with the stated
adjoints.
\epf

The \bfi{linear oscillator group} $LOs(V)$ over $V$ consists of the
elements $[\rho,p,q,\A]$ with invertible $\A$.
One easily checks that the inverse is given by
\lbeq{e.oscInv}
[\rho,p,q,\A]^{-1}=[p^*\A^{-1}q-\rho,-\A^{-*}p,-\A^{-1}q,\A^{-1}],
\eeq
where
\[
\A^{-*}=(\A^{-1})^*=(\A^*)^{-1}.
\]
The \bfi{unitary oscillator group} $UOs(V)$ over $V$ consists
of the unitary elements of $LOs(V)$.

\begin{prop}\label{p.uOs}~\\
(i)  $UOs(V)$ consists of the coherent maps of the form
\lbeq{e.Os}
[\alpha,q,\A]:=[\shalf(i\alpha-q^*q),-\A^*q,q,\A]
\eeq
with unitary $\A\in\Lin V$, $q\in V$, and $\alpha\in\Rz$.

(ii) Product, inverse, and adjoint of unitary elements are given
by
\lbeq{e.uProd}
[\alpha,q,\A][\alpha',q',\A']
=[\alpha+\alpha'-2\im q^*\A q',q'+\A q,\A\A'],
\eeq
\lbeq{e.uInv}
[\alpha,q,\A]^{-1}=[\alpha,q,\A]^*=[-\alpha,-\A^{-1}q,\A].
\eeq
Moreover,
\lbeq{e.BUOs}
[\B][\alpha,q,\A][\B']=[\alpha,\B q,\B\A\B'].
\eeq
\end{prop}

\bepf
(i) Equating \gzit{e.oscAdj} and \gzit{e.oscInv} gives the unitarity
conditions
\[
\ol\rho=p^*\A^{-1}q-\rho,~~~
q=-\A^{-*}p,~~~p=-\A^{-1}q,~~~
\A^*=\A^{-1}.
\]
Thus $\A$ must be unitary and $p=-\A^{-1}q=-\A^*q$. In this case,
$q=-\A^{-*}p$ and
\[
p^*\A^{-1}q=-q^*\A^{-*}\A^{-1}q=-q^*q,
\]
hence the unitarity conditions reduce to $\ol\rho=-q^*q-\rho$, i.e.,
$2\re \rho=-q^*q$. Writing $\alpha=2\im\rho$, (i) follows.

(ii) \gzit{e.uInv} follows from the preceding using \gzit{e.oscAdj},
and \gzit{e.BUOs} follows from \gzit{e.BOs}. 
To obtain the multiplication law we note that
{\small
\[
\bary{ll}
[\alpha,q,\A][\alpha',q',\A']
&=[\shalf(i\alpha-q^*q),-\A^*q,q,\A]
   [\shalf(i\alpha'-q'^*q'),-\A'^*q',q',\A']\\[3mm]
&= [\shalf(i\alpha'-q'^*q')+\shalf(i\alpha-q^*q)-q^*\A q',
     -\A'^*\A^*q -\A'^*q',q+\A q',\A\A']\\[3mm]
&=[\alpha+\alpha'-2\im q^*\A q',q'+\A q,\A\A'].
\eary
\]
} 
Indeed, since $\A^*\A=1$, we have
$-\A'^*\A^*q-\A'^*q'=-(\A\A')^*(q+\A q')$ and
\[
\shalf(i\alpha'-q'^*q')+\shalf(i\alpha-q^*q)-q^*\A q'
=\shalf\Big(i\beta-(q+\A q')^*(q+\A q')\Big),
\]
where
\[
\bary{lll}
i\beta&:=&i\alpha'-q'^*q'+i\alpha-q^*q-2q^*\A q'+(q+\A q')^*(q+\A q')\\
&=&i(\alpha+\alpha')-q^*\A q'+q'^*\A^*q
=i(\alpha+\alpha'-2\im q^*\A q').
\eary
\]
\epf

The unitary coherent maps of the form
\lbeq{e.Hei}
W_\alpha(q):=[\alpha,q,1]~~~(q\in V,\alpha\in\Rz)
\eeq
form the \bfi{Heisenberg group} $H(V)$ over $V$.
The $n$-dimensional \bfi{Weyl group} is the subgroup of $H(\Cz^n)$
consisting of the $W_\alpha(q)$ with real $q$ and $\alpha$.

\begin{prop}
With the symplectic form
\lbeq{e.sigma}
\sigma(q,q'):=-2\im q^*q',
\eeq
we have
\[
W_\alpha(q)W_{\alpha'}(q')=W_{\alpha+\alpha'+\sigma(q,q')}(q+q'),
\]
\[
W_\alpha(q)^{-1}=W_\alpha(q)^*=W_{-\alpha}(-q),
\]
\[
[\B]W_\alpha(q)[\B]^{-1}=W_\alpha(\B q) \iif \B \mbox{ is invertible}.
\]
\end{prop}

\bepf
Specialize Proposition \ref{p.uOs}.
\epf

By \gzit{e.oscCoh}, the action of the Heisenberg group on $Z$ is given 
by
\[
W_\alpha(q)[z_0,\z]=[\shalf(i\alpha-q^*q),-q,q,1][z_0,\z]
=[\shalf(i\alpha-q^*q)+z_0-q^*z,q+\z].
\]

\subsection{Bosonic Fock spaces}\label{ss.bFock}

A \bfi{bosonic Fock space} is a quantum space of a Klauder space
$Kl(V)$.
The quantization map on a Klauder space defines on the corresponding
Fock spaces both a representation of the linear oscillator group
and a unitary representation of the unitary oscillator group $UOs(V)$.
The quantization of the coherent maps in the linear oscillator
semigroup leads to linear operators
$\Gamma([\rho,p^T,q,A])\in\Lin \Qz(Z)$.
Since Klauder spaces are slender, additional linear operators
$\in\Linx\Qz(Z)$ come from the quantization of homogeneous kernels.

In the following, we work with an arbitrary quantum space $\Qz(Z)$,
to demonstrate that everything of interest follows on this level,
without any need to use any explicit integration.
However, to connect to tradition, we note that for $V=\Cz^n$, an
explicit completed quantum space is the space $L^2(\Rz^n,\mu)$ of
square integrable functions of $\Rz$ with respect to the measure $\mu$
given by $d\mu(x)=(2\pi)^{-n/2}e^{-\half x^Tx}dx$ and the inner product
\[
f^*g:=\int d\mu(x)\ol{f(x)}g(x).
\]
To check this we show that the functions
\[
f_z(x):=e^{z_0-\half(x-\z)^2}
\]
constitute the coherent states of finite-dimensional Klauder spaces.
Indeed, using definition \gzit{e.Kosc}, we have
\[
f_z^*f_{z'}
= \int d\mu(x)e^{\ol z_0-\frac{1}{2}(x-\ol \z)^2+z'_0-\half(x-\z')^2}
=e^{\ol z_0+z'_0+\z^*\z'}
 \int d\mu(x)e^{-\half(x-\ol \z)^2-\half(x-\z')^2-\z^*\z'}.
\]
Expanding into powers of $x$ and using the Gaussian integration formula
\[
\int \frac{dx}{(2\pi)^{n/2}}e^{-\half(x-u)^*(x-u)}=1 \for u\in\Cz^n
\]
with $u=\ol \z+\z'$, the last integral can be evaluated to 1, hence
$f_z^*f_{z'}=K(z,z')$. This proves that $K$ is a coherent product and
the $f_z$ are a corresponding family of coherent states.

In the special case $n=1$, we find for
$z=[i\omega\tau-\half\omega^2,\tau+i\omega]$ that
\[
f_z(t)=e^{i\omega\tau-\half\omega^2-\half(t-\tau-i\omega)^2}
=e^{i\omega t}e^{-\half(t-\tau)^2}
\]
is the time-frequency shift by $(\tau,\omega)\in\Rz^2$ of the standard
Gaussian $e^{-\half t^2}$. Thus the general coherent state is a scaled
time-frequency shifted standard Gaussian.

In any quantum space $\Qz(Z)$ of a Klauder space, we write
\[
|\z\>:=|[0,\z]\>
\]
and find from \gzit{e.Kosc} that
\lbeq{e.glauberIP}
\<\z|\z'\>=e^{\z^*\z'},
\eeq
\lbeq{e.zNorm}
|z\>=e^{z_0}|\z\>.
\eeq
Because of \gzit{e.zNorm}, the coherent subspace $Z_0$ consisting of
the $[0,\z]$ with $\z\in V$ has the same quantum space as $Kl(V)$.
We call the coherent spaces $Z_0$ \bfi{Glauber spaces} since the
associated coherent states (originally due to \sca{Schr\"odinger}
\cite{Schr}) were made prominent in quantum optics by
\sca{Glauber} \cite{Gla}. Glauber spaces give a more parsimonious
coherent description of the corresponding Fock space, but Klauder
spaces are much more versatile since they have a much bigger symmetry
group, with corresponding advantages in the applications.

\subsection{Lowering and raising operators}\label{ss.lowering}

In the quantum space of a Klauder space, we introduce an abstract
\bfi{lowering symbol} $a$ and its formal adjoint, the abstract
\bfi{raising symbol} $a^*$.
For $f:V\to\Cz$, we define the formal functions $f(a)$ and $f(a^*)$ to 
be the operators defined by 
\[
f(a):=\a(\wt f),~~~f(a^*):=\a^*(\wt f),
\]
where $\a$ and $\a^*$ are given by \gzit{e.am} and \gzit{e.asm} and 
$\wt f:Z\to \Cz$ is the homogeneous map defined by
\[
\wt f(z):=f(\z) \for z\in Z. 
\]
Clearly $\ol{\wt f}=\wt{\ol f}$, hence \gzit{e.asm} gives 
$f(a)^*=\a(\wt f)^*=\a(\ol{\wt f})=\a(\wt{\ol f})=\ol f(a^*)$, so that
\[
f(a)^*=\ol f(a^*).
\]
For any map $F:V\times V\to\Cz$, we define the homogeneous kernel
$\wt F:Z\times Z\to \Cz$ with $\wt F(z,z'):=F(\z,\z')$, and put
%
\def\:{{{:}}}
\[
\:F(a^*,a)\:~:=N(\wt F)\in\Linx\Qz(Z).
\]
(The pair of colons is the conventional notation for normal ordering
in quantum field theory \cite{DerG}.)

\begin{thm}\label{t.normalOrd}
Let $Z=Kl(V)$. Then:

(i) Every linear operator $A\in\Linx\Qz(Z)$ can be written uniquely in
normally ordered form $A=\:F(a^*,a)\:$.

(ii) The map $F\to \:F\:$ is linear, with $\:1\:=1$ and
\[
\:f(a)^*F(a^*,a)g(a)\:~=~f(a^*)~\:F(a^*,a)\:~g(a);
\]
in particular,
\lbeq{e.fga}
\:f(a)^*g(a)\:~=~f(a)^*g(a).
\eeq
(iii) The quantized coherent maps satisfy
\lbeq{e.genFun}
\Gamma(A)~=~\:e^{\rho+p^*a+a^*q+a^*(\A-1)a}\:
\for A=[\rho,p,q,\A]\in Os(V).
\eeq
(iv) We have the \bfi{Weyl relations}
\[
e^{p^*a}e^{a^*q}=e^{p^*q}e^{a^*q}e^{p^*a}
\]
and the \bfi{canonical commutation relations}
\lbeq{e.ccr0}
(p^*a)(q^*a)=q^*(a)(p^*a),~~~(a^*p)(a^*q)=(a^*q)(a^*p),
\eeq
\lbeq{e.ccr}
(p^*a)(a^*q)-(a^*q)(p^*a)=\sigma(p,q)
\eeq
hold, with the symplectic form \gzit{e.sigma}.
\end{thm}

\bepf
(i) follows from Proposition \ref{p.nowhere} since the coherent product
vanishes nowhere and homogeneous kernels are independent of $z_0$ and
$z_0'$.

(ii) Let $F,G:V\times V\to\Cz$. We then have
\[
\:F+G\:=N(\wt{F+G})=N(\wt{F}+\wt{G})=N(\wt{F})+N(\wt{G})=\:F\:+\:G\:,
\]
\[
\:cF\:=N(\wt{cF})=N(c\wt{F})=cN(\wt{F})=c\:F\:,
\]
which implies that the map $F\to \:F\:$ is linear.

(iii) holds since \gzit{e.Kosc} implies
\[
\<z|A|z'\>=\<z|[\rho,p^*,q,\A]|z'\>
=e^{\ol z_0+\rho+z_0'+p^*\z'+\z^*(q+\A\z')}=X(z,z')K(z,z')
\]
with $X(z,z'):=e^{\rho+p^*\z'+\z^*q+\z^*(\A-1)\z'}$.

(iv) \gzit{e.ccr0} follows directly from the definition of the $f(a)$ 
and $f(a^*)$. The Weyl relations follow from
\[
\bary{lll}
e^{p^*a}e^{a^*q}&=&\:e^{p^*a}\:\,\:e^{a^*q}\:
=\Gamma([0,p,0,1])\Gamma([0,0,q,1])=\Gamma([p^*q,p,q,1])\\
&=&\:e^{p^*q+p^*a+a^*q}\:=\:e^{p^*q}e^{a^*q}e^{p^*a}\:
=e^{p^*q}e^{a^*q}e^{p^*a}.
\eary
\]
Here the first and the last equality are due to \gzit{e.fga}.
The canonical commutation relations \gzit{e.ccr} are obtained from the 
Weyl relations by replacing $p$ and $q$ by $\eps p$ and $\eps q$ with 
$\eps>0$, expanding their exponentials to second order in $\eps$, and 
comparing the coefficients of $\eps^2$.
\epf

If $V=\Cz^n$ we define
\[
a_k:=e_k(a),~~~a_k^*:=e_k(a^*),
\]
where $e_k$ maps $\z$ to $\z_k$. Thus formally, $a$ is a symbolic column
vector with $n$ symbolic \bfi{lowering operators} $a_k$, also called
\bfi{annihilation operators}. Similarly, $a^*$ is a symbolic row vector
with $n$ symbolic \bfi{raising operators} $a_k$, also called
\bfi{creation operators}. They satisfy the standard canonical
commutation relations
\[
a_ja_k=a_ka_j,~~~a_j^*a_k^*=a_k^*a_j^*,
\]
\[
a_ja_k^*-a_k^*a_j=\delta_{jk}
\]
following from \gzit{e.ccr0} and \gzit{e.ccr}.

\subsection{Oscillator algebras}\label{ss.oscAlg}

In the applications one often needs formulas for the Lie products in
the Lie algebras generating the groups discussed in this section, and 
formulas for the resulting representations on Fock space.

The \bfi{oscillator algebra} over $V$ is the Lie algebra $os(V)$ of 
infinitesimal transformations 
\[
X_{\rho,p,q,\A}
:=\lim_{\eps\to 0} \frac{[\eps\rho,\eps p,\eps q,1+\eps\A]-1}{\eps}
=\pmatrix{0 & p^* & \rho \cr 0 & \A & q \cr 0 & 0 & 0}
\]
of $Os(V)$. From \gzit{e.oscProd} we find
\[
X_{\rho,p,q,\A}X_{\rho',p',q',\A'}=[p^*q',\A'^*p,\A q',\A\A'],
\]
whence the Lie product is
\lbeq{e.osLie}
[X_{\rho,p,q,\A},X_{\rho',p',q',\A'}]
=[p^*q'-p'^*q,\A'^*p-\A^*p',\A q'-\A'q,\A\A'-\A'\A].
\eeq
The corresponding representation on Fock space follows from 
\gzit{e.genFun} and is given by
\lbeq{e.dGammaOs}
d\Gamma(X_{\rho,p,q,\A}):=\lim_{\eps\to 0} 
               \frac{\Gamma([\eps\rho,\eps p,\eps q,1+\eps\A])-1}{\eps}
=\rho+p^*a+a^*q+a^*\A a.
\eeq
The \bfi{unitary oscillator algebra} over $V$ is the Lie algebra 
$uos(V)$ of infinitesimal transformations 
\[
X_{\alpha,q,\A}:=\lim_{\eps\to 0}
                 \frac{[\eps\alpha,\eps q,1+\eps\A]-1}{\eps}
=X_{\half i\alpha,-q,q,\A}
\]
of $UOs(V)$. By specializing \gzit{e.osLie}, the Lie product is found to
be 
\lbeq{e.uosLie}
[X_{\alpha,q,\A},X_{\alpha',q',\A'}]
=X_{i\sigma(q,q'),\A q'-\A'q,\A\A'-\A'\A}.
\eeq
The corresponding representation on Fock space follows from 
\gzit{e.dGammaOs} and is given by
\lbeq{e.dGammaUOs}
d\Gamma(X_{\alpha,q,\A})
:=\lim_{\eps\to 0} \frac{\Gamma([\eps\alpha,\eps q,1+\eps\A])-1}{\eps}
=\shalf i\alpha-q^*a+a^*q+a^*\A a.
\eeq
The \bfi{Heisenberg algebra} over $V$ is the Lie algebra $h(V)$
of infinitesimal transformations 
\[
X_{\alpha,q}
:=\lim_{\eps\to 0} \frac{W_{\eps\alpha}(\eps q)-1}{\eps}
=X_{\alpha,q,0}
\]
of $H(V)$. By specializing \gzit{e.uosLie}, the Lie product is found 
to
be 
\lbeq{e.heiLie}
[X_{\alpha,q},X_{\alpha',q'}]=X_{\sigma(q,q'),0}.
\eeq
The corresponding representation on Fock space follows from 
\gzit{e.dGammaUOs} and is given by
\lbeq{e.dGammaHei}
d\Gamma(X_{\alpha,q})
:=\lim_{\eps\to 0} \frac{\Gamma(W_{\eps\alpha}(\eps q))-1}{\eps}
=\shalf i\alpha-q^*a+a^*q.
\eeq

\section{Declarations}

\subsection{Corresponding Author} 
Arnold Neumaier

\subsection{Funding} 
No funding was received for conducting this study.

\subsection{Conflicts of interest/Competing interests}
The authors have no financial or proprietary interests in any material 
discussed in this article.

\subsection{Availability of data and material (data transparency)}
Not applicable

\subsection{Code availability}
Not applicable


\bigskip

\end{document}